\documentclass[english]{article}
\usepackage{ae,aecompl}

\usepackage[T1]{fontenc}
\usepackage[latin9]{inputenc}
\usepackage{geometry}
\usepackage{units}
\usepackage{amsmath}
\usepackage{stackrel}
\usepackage{graphicx}

\makeatletter
\numberwithin{equation}{section}

\makeatother

\usepackage{babel}
\begin{document}
\title{A Function Based on Chebyshev Polynomials as an\\
Alternative to the Sinc Function in FIR Filter Design}
\author{Paul W. Oxby}
\maketitle
\begin{abstract}
The Fourier transform of the sinc function is a step function, the
ideal frequency response for a low pass filter. For this reason a
truncated sinc function is often used as the basis for the design
of discrete linear-phase FIR filters. However the Fourier transform
of a truncated sinc function exhibits ripple in the pass band due
to the Gibbs phenomenon. This paper introduces an alternative design
function, expressible in terms of Chebyshev polynomials, whose Fourier
transform decreases monotonically in the pass band. Furthermore this
function features an intrinsic window function with an adjustable
parameter influencing the Fourier transform in the transition and
stop bands.\smallskip{}

While the Fourier transform of the alternative filter function doesn't
exhibit ripple in the pass band, the FIR filter based on evaluating
the function at discrete intervals can exhibit ripple in the pass
band. However the amplitude of this ripple approaches zero as the
number of the intervals increases. This suggests that the pass band
ripple is due to the discretization of the filter function. An algorithm
is presented that compensates for the effect of discretization giving
a discrete FIR filter with an intrinsic window function and whose
frequency response is free of ripple in the pass band.
\end{abstract}

\section*{\qquad{}\qquad{}\qquad{}\qquad{}\qquad{}\qquad{}Table of Contents}

\[
\begin{array}{lr}
1\;\;\mathrm{The\:Grace\:Polynomials}\quad.......................................................................... & 1\\
2\;\;\mathrm{The\:Limiting\:Case\:of\:the\:Grace\:Polynomials\:as\:}n\rightarrow\infty\quad......................... & 4\\
3\;\;\mathrm{The\:Grace\:Function\:and\:Its\:Fourier\:Transform\quad}...................................... & 5\\
4\;\;\mathrm{The\:Scale\:Factor\:of\:the\:Grace\:Function\:Fourier\:Transform}\quad.................... & 6\\
5\;\;\mathrm{The\:Taylor\:Series\:of\:the\:Grace\:Function\:Fourier\:Transform}\quad................... & 8\\
6\;\;\mathrm{The\:Monotonicity\:of\:the\:Grace\:Function\:Fourier\:Transform}\quad.................. & 11\\
7\;\;\mathrm{The\:Grace\:Filter\:and\:Its\:Properties\quad........................................................} & 15\\
8\;\;\mathrm{The\:Discretization\:Compensation\:Algorithm}\quad......................................... & 18\\
9\;\;\mathrm{The\:Performance\:of\:the\:Adjusted\:Grace\:Filter}\quad....................................... & 22\\
\mathrm{References}\quad................................................................................................. & 24\\
\mathrm{Acknowledgments}\quad...................................................................................... & 24\\
\mathrm{Contact}\quad..................................................................................................... & 24\\
\mathrm{Appendix\:1:The\:Grace\:Filter\:Coefficients\:Program}\quad.................................. & 24\\
\mathrm{Appendix\:2:The\:Adjusted\:Grace\:Filter\:Program\quad}...................................... & 25\\
\mathrm{Appendix\:3:More\:on\:the\:Grace\:Function\:Fourier\:Transform}\quad..................... & 26
\end{array}
\]

\section{The Grace Polynomials}

The coefficients of the Savitzky-Golay smoothing filter are derived
from a least-squares polynomial fit to a hypothetical set of equally
spaced data {[}1{]}. The frequency response of this filter decreases
monotonically in the pass band so there is no ripple in the pass band.
However the frequency response of the filter in the stop band is relatively
poor {[}2{]}. An analysis of the properties of the Savitzky-Golay
filter has lead to the development of an alternative to the sinc function
in the design of linear phase FIR digital filters.\newpage{}

This alternative to the sinc function, the Grace polynomials, $\mathrm{Gp}(x,n)$,
can be expressed in several equivalent forms for \emph{x} on the interval
$\left[-1,1\right]$. $\mathrm{Gp}(x,n)$ is expressed in terms of
Chebyshev polynomials of either the first kind or the second kind
as: 

\begin{equation}
\mathrm{Gp}(x,n)=(-1)^{n}\:\frac{T_{2n+1}(x)-T_{2n-1}(x)}{4nx}
\end{equation}

\begin{equation}
\mathrm{Gp}(x,n)=(-1)^{n}\left(1-x^{2}\right)\dfrac{-U_{2n-1}(x)}{2nx}
\end{equation}

The odd-indexed Chebyshev polynomials, being odd functions, have a
factor of \emph{x} which cancels the \emph{x} in the denominators
of Equations 1.1 and 1.2. This makes $\mathrm{Gp}(x,n)$ an even polynomial
of degree $2n$. The following recurrence relation for $\mathrm{Gp}(x,n)$
avoids the problematic division by \emph{x}:

\begin{equation}
\begin{array}{l}
u_{0}=1\\
v_{0}=0\\
\\
u_{i}=2v_{i-1}\,x^{2}-u_{i-1}\quad(\mathrm{for}\:i=1,...\,n)\\
v_{i}=2u_{i}-v_{i-1}\\
\\
\mathrm{Gp}(x,n)=(-1)^{n}\left(1-x^{2}\right)\dfrac{v_{n}}{2n}
\end{array}
\end{equation}

Applying the trigonometric expressions for Chebyshev polynomials gives:

\begin{equation}
\mathrm{Gp}(x,n)=(-1)^{n}\:\frac{-\sin\left(2n\arccos x\right)\sqrt{1-x^{2}}}{2nx}
\end{equation}

The 2\emph{n} roots of $\mathrm{Gp}(x,n)$ are given by the 2\emph{n}
roots of the term $\sin(2n\arccos x)$ on the interval $\left[-1,1\right]$
: 

\begin{equation}
\mathrm{Gp}\left(\pm\sin\left(\dfrac{\pi}{2}\,\dfrac{i}{n}\right),n\right)=0\quad(\mathrm{for}\:i=1,...\,n)
\end{equation}

Because the roots are symmetric the polynomial $\mathrm{Gp}(x,n)$
can be expressed as a product of \emph{n} binomials:

\begin{equation}
\mathrm{Gp}(x,n)=\dfrac{1}{4n}\stackrel[i=1]{n}{\prod}4\left[\sin^{2}\left(\dfrac{\pi}{2}\,\dfrac{i}{n}\right)-x^{2}\right]
\end{equation}

$\mathrm{Gp}(x,n)$ can also be expressed as a power series in $x^{2}$:

\begin{equation}
\mathrm{Gp}(x,n)=\frac{1}{n}\stackrel[i=0]{n}{\sum}G_{n,i}\,x^{2\,i}
\end{equation}

where the coefficients, $G_{n,i}$, of the power series are given
by:

\begin{equation}
G_{n,i}=(-4)^{i}\,\dfrac{\left(2n^{2}+i\right)\left(n+i-1\right)!}{2\left(2i+1\right)!\,\left(n-i\right)!}\quad(\mathrm{for}\:i=0,...\,n)
\end{equation}

These coefficients may be generated row-wise by the recurrence relation:

\begin{equation}
\begin{array}{l}
G_{n,0}=n\\
G_{n,i}=-G_{n,i-1}\,\dfrac{2\left(n^{2}-(i-1)^{2}\right)(2n^{2}+i)}{i\left(2i+1\right)\left(2n^{2}+i-1\right)}\quad(\mathrm{for}\:i=1,...\,n)
\end{array}
\end{equation}

\newpage{}

The values of $G_{n,i}$ for the first nine values of $\mathit{n}$
are given in Table 1.1:
\begin{center}
Table 1.1: Coefficients $G_{n,i}$ for \emph{n} ranging from 1 to
9.
\[
\begin{array}{rrrrrrrrrrr}
 & \;i\quad0 & 1 & 2 & 3 & 4 & 5 & 6 & 7 & 8 & 9\\
n\\
1 & 1 & -1\\
2 & 2 & -6 & 4\\
3 & 3 & -19 & 32 & -16\\
4 & 4 & -44 & 136 & -160 & 64\\
5 & 5 & -85 & 416 & -848 & 768 & -256\\
6 & 6 & -146 & 1036 & -3200 & 4864 & -3584 & 1024\\
7 & 7 & -231 & 2240 & -9696 & 21760 & -26368 & 16384 & -4096\\
8 & 8 & -344 & 4368 & -25152 & 77440 & -136192 & 137216 & -73728 & 16384\\
9 & 9 & -489 & 7872 & -58080 & 233728 & -555776 & 802816 & -692224 & 327680 & -65536
\end{array}
\]
\par\end{center}

The values of $G_{n,i}$ also satisfy the following recurrence relation
for $n>1$:

\begin{equation}
\begin{array}{l}
G_{n,0}=n\\
G_{n,i}=2\,G_{n-1,i}-4\,G_{n-1,i-1}-G_{n-2,i}\quad(\mathrm{for}\:n>1,\:i=1,...\,n)
\end{array}
\end{equation}
\smallskip{}

The second derivative of $\mathrm{Gp}(x,n)$ at $x=0$ is $-(4n^{2}+2)/3$.
The sinc function can be scaled to give it the same second derivative
at $x=0$:
\begin{equation}
\mathrm{S}(x,n)=\mathrm{sinc}\left(x\:\sqrt{4n^{2}+2}\right)
\end{equation}

$\mathrm{S}(x,n)$ and $\mathrm{Gp}(x,n)$ are plotted for $n=12$
in Figure 1.1:
\begin{center}
Figure 1.1: $\mathrm{S}(x,n)$ and $\mathrm{Gp}(x,n)$ for $n=12$
\par\end{center}

\begin{center}
\includegraphics[scale=0.6]{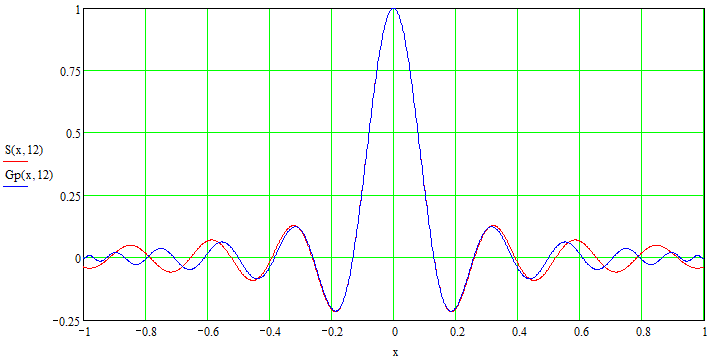}
\par\end{center}

The functions $\mathrm{Gp}(x,12)$ and $\mathrm{S}(x,12)$ in Figure
1.1 are almost coincident near $x=0$ by design. However they are
also close to being coincident in the first two sidelobes and this
is not a coincidence. In the next section it will be shown that the
functions $\mathrm{Gp}(x,n)$ and $\mathrm{S}(x,n)$ converge as $n\rightarrow\infty$.\newpage{}

\section{The Limiting Case of the Grace Polynomials as $n\rightarrow\infty$}

In order to show that the functions $\mathrm{Gp}(x,n)$ and $\mathrm{S}(x,n)$
converge as $n\rightarrow\infty$ the scaling of the sinc function
illustrated in Figure 1.1 will be modified. Instead of scaling the
sinc function to align with a specific Grace polynomial at the origin,
the Grace polynomials will be scaled to align with the sinc function
at the origin. This scaling is done by a change of variable from $x$
in Equation 1.7 to $x^{\prime}$ as given by the following linear
transformation:

\begin{equation}
x^{\prime}=x\:\sqrt{4n^{2}+2}
\end{equation}

This transformation simply rescales the x axis in Figure 1.1. The
rescaled Grace polynomial will be denoted by $\mathrm{Gp^{\prime}}(x^{\prime},n)$
and by incorporating the $\nicefrac{1}{n}$ term of Equation 1.7 into
the summation term $\mathrm{Gp^{\prime}}(x^{\prime},n)$ can be expressed
as this power series in $x^{\prime\,2}$:

\begin{equation}
\mathrm{Gp^{\prime}}(x^{\prime},n)=\stackrel[i=0]{n}{\sum}G_{n,i}^{\prime}\,x^{\prime\,2\,i}
\end{equation}

The scale factor of $\sqrt{4n^{2}+2}$ makes the second derivative
of $\mathrm{Gp^{\prime}}(x^{\prime},n)$ evaluated at $x^{\prime}=0$
equal to the second derivative of the sinc function evaluated at zero
which is $-\nicefrac{1}{3}$. Note that the same scale factor appears
in Equation 1.11 for the same purpose.\smallskip{}

A recurrence relation for the coefficients $G_{n,i}^{\prime}$ is
given by incorporating the scale factor $\sqrt{4n^{2}+2}$ of Equation
2.1 into the recurrence relation of Equation 1.9:

\begin{equation}
\begin{array}{l}
G_{n,0}^{\prime}=1\\
G_{n,i}^{\prime}=-G_{n,i-1}^{\prime}\,\dfrac{2\left(n^{2}-(i-1)^{2}\right)\left(2n^{2}+i\right)}{i\left(2i+1\right)\left(2n^{2}+i-1\right)\left(4n^{2}+2\right)}\quad(\mathrm{for}\:i=1,...\,n)
\end{array}
\end{equation}

In this equation the scale factor is squared because the power series
for $\mathrm{Gp^{\prime}}(x^{\prime},n)$ is in $x^{\prime}$ squared.
For any fixed value of\emph{ i} the following limit applies:

\begin{equation}
\underset{n\rightarrow\infty}{\lim}\:\frac{G_{n,i}^{\prime}}{G_{n,i-1}^{\prime}}=\frac{-1}{2i\left(2i+1\right)}
\end{equation}

Given that $G_{n,0}^{\prime}=1$, the limiting value of $G_{n,i}^{\prime}$
as $n\rightarrow\infty$ is given by: 

\begin{equation}
\underset{n\rightarrow\infty}{\lim}G_{n,i}^{\prime}=\frac{-\left(-1\right)^{i}}{\left(2i+1\right)!}\quad(\mathrm{for}\:i=0,...\,n)
\end{equation}

Note that while the range of \emph{x} is $\left[-1,1\right]$ the
limiting range of $x^{\prime}$ as $n\rightarrow\infty$ is $\left[-\infty,\infty\right]$.
Substituting Equation 2.5 into Equation 2.2 gives the desired result:

\begin{equation}
\begin{array}{c}
\underset{n\rightarrow\infty}{\lim}\mathrm{Gp^{\prime}}(x^{\prime},n)=1-\dfrac{x^{\prime\,2}}{3!}+\dfrac{x^{\prime\,4}}{5!}-\dfrac{x^{\prime\,6}}{7!}+\dfrac{x^{\prime\,8}}{9!}-\cdots\\
\\
=\mathrm{sinc}(x^{\prime})\quad\mathrm{Q.E.D.}
\end{array}
\end{equation}

This result shows that $\mathrm{Gp}(x,n)$ is closely related to the
sinc function. In a sense the polynomials $\mathrm{Gp}(x,n)$ can
be considered as modifications of the sinc function that gracefully
approach zero at the ends of the interval $\left[-1,1\right]$. Acknowledgment
of this graceful behaviour is the basis for naming the polynomials
$\mathrm{Gp}(x,n)$ the Grace polynomials.\newpage{}

\section{The Grace Function and Its Fourier Transform}

It follows from Equations 1.2, 1.3 or 1.6 that $1-x^{2}$ is a factor
of the Grace polynomials $\mathrm{Gp}(x,n)$. These polynomials can
be generalized by raising the factor $1-x^{2}$ to a power, giving
a parametric generalization of the Grace polynomials, the Grace function:

\begin{equation}
\mathrm{G}(x,n,p)=\mathrm{Gp}(x,n)\left(1-x^{2}\right)^{\,p-\nicefrac{1}{2}}\quad(\mathrm{for}\:p=0,\,1,\,2,\,...)
\end{equation}

The exponent is expressed as $p-\nicefrac{1}{2}$ because the values
of \emph{p} are to be constrained to nonnegative integers. As $\mathit{p}$
increases the function $\left(1-x^{2}\right)^{p-\nicefrac{1}{2}}$
approaches a Gaussian function with a standard deviation of $(2p)^{\nicefrac{-1}{2}}$.
In effect, \emph{p} parameterizes a window function. But unlike the
ad hoc window functions commonly applied to the sinc function, it
will be shown that the term $(1-x^{2})^{p-\nicefrac{1}{2}}$ can be
considered as a window function intrinsic to the Grace function.\smallskip{}

The scaled Fourier transform, $\mathrm{g}(\phi,n,p)$, of $\mathrm{G}(x,n,p)$
can be written as:

\begin{equation}
\mathrm{g}(\phi,n,p)=\frac{1}{\mathrm{a}(n,p)}\int_{-1}^{1}\mathrm{G}(x,n,p)\,\cos\left[\mathrm{b}(n,p)\,\pi\,\phi\,x\right]\,dx
\end{equation}

The scale factors $\mathrm{a}(n,p)$ and $\mathrm{b}(n,p)$ of Equation
3.2 are determined by imposing the following conditions:

\begin{equation}
\mathrm{\mathrm{g}}(0,n,p)=1
\end{equation}

\begin{equation}
\int_{0}^{\infty}\mathrm{\mathrm{g}}(\phi,n,p)^{2}d\phi=1
\end{equation}

With these two conditions the scale factors $\mathrm{a}(n,p)$ and
$\mathrm{b}(n,p)$ are given by:

\begin{equation}
\mathrm{a}(n,p)=\int_{-1}^{1}\mathrm{G}\left(x,n,p\right)dx
\end{equation}

\begin{equation}
\mathrm{b}(n,p)=\frac{1}{\mathrm{a}(n,p)^{2}}\int_{-1}^{1}\mathrm{G}(x,n,p)^{2}dx
\end{equation}

The condition of Equation 3.4 is imposed so that the value $\phi=1$
will roughly correspond to the cutoff frequency, $\phi_{c}$, for
which $\mathrm{g}(\phi_{c},n,p)^{2}=\nicefrac{1}{2}$. The following
inequality places the value $\mathrm{g}(1,n,p)^{2}$ in the transition
band of the frequency response for all values of \emph{n} and $p<n$:

\begin{equation}
\mathrm{0.45<}\:\mathrm{g}(1,n,p)^{2}<0.55\quad(\mathrm{for}\:p<n)
\end{equation}

Therefore when considering the properties of the Fourier transform
of the Grace function in the pass band it is sufficient to consider
the Fourier transform for values of $\phi$ on the interval $\left[0,1\right]$.\newpage The
Fourier transforms, $\mathrm{g}(\phi,n,p)$ and $\mathrm{s}(\phi,n)$,
of $\mathrm{G}(x,n,p)$ and $\mathrm{S}(x,n)$ for $n=12$ and $p=0$
are plotted in Figure 3.1:
\begin{center}
Figure 3.1: The Fourier transforms $\mathrm{g}(\phi,n,p)$ and $\mathrm{s}(\phi,n)$
for $n=12\,\mathrm{and}\,p=0$
\par\end{center}

\begin{center}
\includegraphics[scale=0.6]{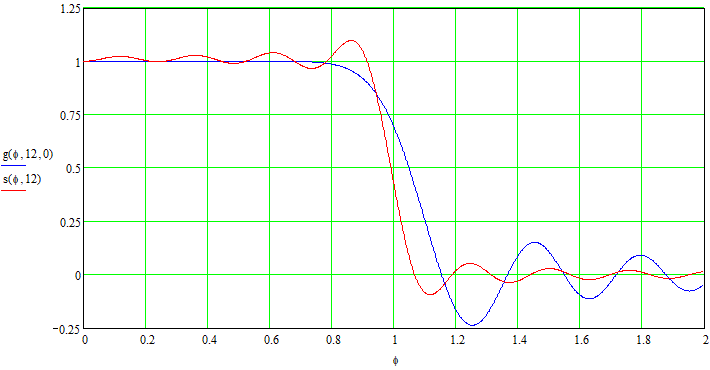}
\par\end{center}

The range of $\phi$ is $\left[0,\infty\right]$ but it is the pass
band of the Fourier transform that is of primary interest. The truncation
of the sinc function causes ripple in the pass band of the transform
due to the Gibbs phenomenon. The transform of the Grace function decreases
monotonically in the pass band. The price to be paid for relative
flatness in the pass band is poorer response relative to the sinc
function in the transition and stop bands. However the case where
$p=0$ is the worst case for the stop band response of the Grace function.
The ripple in the stop band significantly decreases with increasing
values of \emph{p} (see Section 7).

\section{The Scale Factor of the Grace Function Fourier Transform}

Equation 3.5 for $\mathrm{a}(n,p)$ can be written as:

\begin{equation}
\mathrm{a}(n,p)=\dfrac{1}{n}\int_{-1}^{1}\stackrel[i=0]{n}{\sum}G_{n,i}\,x{}^{2i}\left(1-x^{2}\right)^{\,p-\nicefrac{1}{2}}dx
\end{equation}

The definite integral can be evaluated analytically using Equation
1.8 and this identity derived from the Euler beta function:

\begin{equation}
\int_{-1}^{1}x^{2u}\left(1-x^{2}\right)^{v-\nicefrac{1}{2}}dx=\frac{\Gamma\left(u+\nicefrac{1}{2}\right)\Gamma\left(v+\nicefrac{1}{2}\right)}{\Gamma\left(u+v+1\right)}=\dfrac{\pi\,(2u)!\,(2v)!}{4^{u+v}\,u!\,v!\,(u+v)!}
\end{equation}

where \emph{u} and \emph{v} are nonnegative integers. Combining Equations
1.8, 4.1 and 4.2 gives::

\begin{equation}
\mathrm{a}(n,p)=\dfrac{\pi}{n}\stackrel[i=0]{n}{\sum}\left[(-4)^{i}\dfrac{\left(2n^{2}+i\right)\left(n+i-1\right)!}{\left(2i+1\right)!\left(n-i\right)!}\dfrac{\left(2i\right)!\left(2p\right)!}{4^{i+p}\,i!\,p!\left(i+p\right)!}\right]\quad(\mathrm{for}\:n>0)
\end{equation}

It follows from this that $\mathrm{a}(n,p)$ is a rational multiple
of $\pi$.\newpage{}

The following transformation will be applied to the scale factor $\mathrm{a}(n,p)$:
\begin{equation}
\mathrm{d}(n,p)=4^{p}\left[1-\frac{2n}{\pi}\,\mathrm{a}(n,p)\right]
\end{equation}

For the purpose of initializing the recurrence relation in this section
$\mathrm{a}(n,p)$ will be defined to be zero for $n=0$. With $\mathrm{a}(0,p)\equiv0$
it follows that $\mathrm{d}(0,p)=4^{p}$. The values of $\mathrm{d}(n,p)$
for values of \emph{n} and \emph{p} ranging from 0 to 9 are given
in Table 4.1:
\begin{center}
Table 4.1: Values of the transformed scale factor, $\mathrm{d}(n,p)$.
\[
\begin{array}{rrrrrrrrrrr}
 & p\quad0 & \;\,1 & 2 & 3 & 4 & 5 & 6 & 7 & 8 & 9\\
n\\
0 & 1 & 4 & 16 & 64 & 256 & 1024 & 4096 & 16384 & 65536 & 262144\\
1 & 0 & 1 & 6 & 29 & 130 & 562 & 2380 & 9949 & 41226 & 169766\\
2 & 0 & 0 & 1 & 8 & 46 & 232 & 1093 & 4944 & 21778 & 94184\\
3 & 0 & 0 & 0 & 1 & 10 & 67 & 378 & 1941 & 9402 & 43796\\
4 & 0 & 0 & 0 & 0 & 1 & 12 & 92 & 576 & 3214 & 16664\\
5 & 0 & 0 & 0 & 0 & 0 & 1 & 14 & 121 & 834 & 5036\\
6 & 0 & 0 & 0 & 0 & 0 & 0 & 1 & 16 & 154 & 1160\\
7 & 0 & 0 & 0 & 0 & 0 & 0 & 0 & 1 & 18 & 191\\
8 & 0 & 0 & 0 & 0 & 0 & 0 & 0 & 0 & 1 & 20\\
9 & 0 & 0 & 0 & 0 & 0 & 0 & 0 & 0 & 0 & 1
\end{array}
\]
\par\end{center}

The values of $\mathrm{d}(n,p)$ in this table can be generated column
by column by this recurrence relation:

\begin{equation}
\begin{array}{l}
\mathrm{d}(0,p)=4^{p}\\
\mathrm{d}(n,0)=0\quad(\mathrm{for}\,n>0)\\
\mathrm{d}(n,p)=\mathrm{d}(n-1,p-1)+2\:\mathrm{d}(n,p-1)+\mathrm{d}(n+1,p-1)
\end{array}
\end{equation}

The diagonal patterns in Table 4.1 involving the variables \emph{n}
and \emph{p} reflect the symmetry between \emph{u} and \emph{v} on
the right hand sides of Equation 4.2. This supports regarding the
term $(1-x^{2})^{\,p-\nicefrac{1}{2}}$ in Equation 3.1 as a window
function intrinsic to the Grace function in contrast to the ad hoc
window functions commonly applied to the sinc function. It is a particular
feature of the Grace function that $\mathrm{d}(n,p)=0$ for $n>p$.
The significance of this will become apparent when the properties
of the Fourier transform of the Grace function are examined. \smallskip{}

The pattern of zeroes in Table 4.1 implies the following equation
for $\mathrm{a}(n,p)$ for $p<n$:

\begin{equation}
\mathrm{a}(n,p)=\dfrac{\pi}{2n}\quad(\mathrm{for}\:p<n)
\end{equation}

The value of the scale factor $\mathrm{b}(n,p)$, defined by Equation
3.6, is given for $p<n$ by the following equations:

\begin{equation}
\mathrm{b}(n,p)\:=\begin{cases}
\:\dfrac{n}{\pi^{2}}\stackrel[i=1]{n}{\sum}\dfrac{16}{(4i-1)\,(4i-3)} & \mathrm{for}\:p=0\\
\\
\:\dfrac{2n}{\pi}-\dfrac{1}{\pi^{2}}\stackrel[i=1]{2p}{\prod}\dfrac{2i}{2i-1}+O\left(n^{-4p-2}\right) & \mathrm{for}\:1\leq p<n
\end{cases}
\end{equation}

For $p>0$ and $n\gg1$ the term $O\left(n^{-4p-2}\right)$ is negligible.
A fairly good approximation for $\mathrm{b}(n,p)$ is given by:

\begin{equation}
\mathrm{b}(n,p)\:\approx\:\dfrac{2n}{\pi}-\dfrac{1}{\pi^{2}}\,\sqrt{2\pi p+\dfrac{1+\pi p}{1+4p}}\quad(\mathrm{for}\:p<n)
\end{equation}
\newpage{}

\section{The Taylor Series of the Grace Function Fourier Transform}

The Grace function's Fourier transform, $\mathrm{g}(\phi,n,p)$, is
an even function in $\phi$ therefore its odd order derivatives evaluated
at $\phi=0$ are zero. The \emph{k}th even order derivative of $\mathrm{g}(\phi,n,p)$
evaluated at $\phi=0$ is given by:

\begin{equation}
\left.\frac{d^{2k}\mathrm{g}(\phi,n,p)}{d\phi^{2k}}\right|_{\phi=0}=\left(-1\right)^{k}\frac{\left[\,\pi\,\mathrm{b}(n,p)\,\right]^{2k}}{\mathrm{a}(n,p)}\int_{-1}^{1}\mathrm{G}(x,n,p)\,x^{2k}\,dx\quad(\mathrm{for}\:k=1,\,2,\,3,\,...)
\end{equation}

The \emph{k}th even order derivative of $\mathrm{g}(\phi,n,p)$, evaluated
at $\phi=0$ will be transformed by a multiplication of a function
of \emph{n}, \emph{p} and \emph{k}. The transformed derivatives will
be denoted by $\mathrm{d}(n,p,k)$ which  is given by:

\begin{equation}
\mathrm{d}(n,p,k)=\dfrac{2n}{\pi}\frac{4^{p+k}\,\mathrm{a}(n,p)}{\left[\,\pi\,\mathrm{b}(n,p)\,\right]^{2k}}\left.\dfrac{d^{2k}\mathrm{g}(\phi,n,p)}{d\phi^{2k}}\right|_{\phi=0}=\left(-1\right)^{k}4^{p+k}\,\dfrac{2n}{\pi}\int_{-1}^{1}\mathrm{G}(x,n,p)\,x^{2k}\,dx
\end{equation}

The integral can be evaluated analytically by using Equations 1.8
and 4.2 giving:

\begin{equation}
\mathrm{d}(n,p,k)=-\dfrac{(2\,(p+1))!}{2\,(p+1)!}\stackrel[i=1]{n}{\sum}\left[\dfrac{(2\,(k+i-1))!}{(k+i-1)!}\:\dfrac{(-1)^{k+i}\,(n+i-1)!}{(2i-1)!\:(p+k+i)!\:(n-i)!}\right]
\end{equation}

Equation 5.3 can be efficiently evaluated by calculating the terms
in the summation recursively:

\begin{equation}
\begin{array}{c}
\begin{array}{l}
u_{1}=\dfrac{(2\,(p+1))!}{2\,(p+1)!}\:\dfrac{(2k)!}{k!}\:\dfrac{(-1)^{k}\,n}{(p+k+1)!}\\
\\
u_{i+1}=-u_{i}\dfrac{(n+i)\:(n-i)\:(2k+2i-1)}{i\:(2i+1)\:(p+k+i+1)}\quad(\mathrm{for}\:i=1,...\,n-1)\\
\\
\mathrm{d}(n,p,k)=\stackrel[i=1]{n}{\sum}u_{i}
\end{array}\end{array}
\end{equation}

For the purpose of initializing the recurrence relations in this section
$\mathrm{d}(n,p,k)$ will be defined to be zero for $n=0$, i.e.,
$\mathrm{d}(0,p,k)\equiv0$. Tables 5.1, 5.2, 5.3 and 5.4 give the
values of $\mathrm{d}(n,p,k)$ for $k=1$, 2, 3 and 4 respectively
and for values of \emph{n} and \emph{p} ranging from 0 to 9.
\begin{center}
Table 5.1: $\mathrm{d}(n,p,1)\propto d^{2}\mathrm{g}(\phi,n,p)/d\phi^{2}|_{\phi=0}$
\[
\begin{array}{rrrrrrrrrrr}
 & p\quad0 & 1 & 2 & 3 & 4 & 5 & 6 & 7 & 8 & 9\\
n\\
0 & 0 & 0 & 0 & 0 & 0 & 0 & 0 & 0 & 0 & 0\\
1 & -1 & -2 & -5 & -14 & -42 & -132 & -429 & -1430 & -4862 & -16796\\
2 & 0 & -1 & -4 & -14 & -48 & -165 & -572 & -2002 & -7072 & -25194\\
3 & 0 & 0 & -1 & -6 & -27 & -110 & -429 & -1638 & -6188 & -23256\\
4 & 0 & 0 & 0 & -1 & -8 & -44 & -208 & -910 & -3808 & -15504\\
5 & 0 & 0 & 0 & 0 & -1 & -10 & -65 & -350 & -1700 & -7752\\
6 & 0 & 0 & 0 & 0 & 0 & -1 & -12 & -90 & -544 & -2907\\
7 & 0 & 0 & 0 & 0 & 0 & 0 & -1 & -14 & -119 & -798\\
8 & 0 & 0 & 0 & 0 & 0 & 0 & 0 & -1 & -16 & -152\\
9 & 0 & 0 & 0 & 0 & 0 & 0 & 0 & 0 & -1 & -18
\end{array}
\]
\par\end{center}

The values of $\mathrm{d}(n,p,1)$ in this table can be generated
column by column by this recurrence relation:

\begin{equation}
\begin{array}{l}
\mathrm{d}(0,p,1)=0\\
\mathrm{d}(n,0,1)=0\quad(\mathrm{for}\,n>1)\\
\mathrm{d}(n,p,1)=\mathrm{d}(n-1,p-1,1)+2\:\mathrm{d}(n,p-1,1)+\mathrm{d}(n+1,p-1,1)
\end{array}
\end{equation}

Note that the last line of this recurrence is identical to the last
line of the recurrence of Equation 4.5.
\begin{center}
Table 5.2: $\mathrm{d}(n,p,2)\propto d^{4}\mathrm{g}(\phi,n,p)/d\phi^{4}|_{\phi=0}$
\[
\begin{array}{rrrrrrrrrrr}
 & p\quad0 & 1 & 2 & 3 & 4 & 5 & 6 & 7 & 8 & 9\\
n\\
0 & 0 & 0 & 0 & 0 & 0 & 0 & 0 & 0 & 0 & 0\\
1 & 2 & 3 & 6 & 14 & 36 & 99 & 286 & 858 & 2652 & 8398\\
2 & -1 & 0 & 2 & 8 & 27 & 88 & 286 & 936 & 3094 & 10336\\
3 & 0 & -1 & -2 & -3 & -2 & 11 & 78 & 364 & 1496 & 5814\\
4 & 0 & 0 & -1 & -4 & -12 & -32 & -78 & -168 & -272 & 0\\
5 & 0 & 0 & 0 & -1 & -6 & -25 & -90 & -300 & -952 & -2907\\
6 & 0 & 0 & 0 & 0 & -1 & -8 & -42 & -184 & -731 & -2736\\
7 & 0 & 0 & 0 & 0 & 0 & -1 & -10 & -63 & -322 & -1463\\
8 & 0 & 0 & 0 & 0 & 0 & 0 & -1 & -12 & -88 & -512\\
9 & 0 & 0 & 0 & 0 & 0 & 0 & 0 & -1 & -14 & -117
\end{array}
\]
\par\end{center}

\begin{center}
Table 5.3: $\mathrm{d}(n,p,3)\propto d^{6}\mathrm{g}(\phi,n,p)/d\phi^{6}|_{\phi=0}$
\[
\begin{array}{rrrrrrrrrrr}
 & p\quad0 & 1 & 2 & 3 & 4 & 5 & 6 & 7 & 8 & 9\\
n\\
0 & 0 & 0 & 0 & 0 & 0 & 0 & 0 & 0 & 0 & 0\\
1 & -5 & -6 & -10 & -20 & -45 & -110 & -286 & -780 & -2210 & -6460\\
2 & 4 & 2 & 0 & -5 & -20 & -66 & -208 & -650 & -2040 & -6460\\
3 & -1 & 2 & 5 & 10 & 19 & 34 & 52 & 40 & -170 & -1292\\
4 & 0 & -1 & 0 & 4 & 16 & 50 & 144 & 400 & 1088 & 2907\\
5 & 0 & 0 & -1 & -2 & -1 & 10 & 60 & 248 & 901 & 3078\\
6 & 0 & 0 & 0 & -1 & -4 & -10 & -16 & 5 & 188 & 1102\\
7 & 0 & 0 & 0 & 0 & -1 & -6 & -23 & -70 & -175 & -322\\
8 & 0 & 0 & 0 & 0 & 0 & -1 & -8 & -40 & -160 & -556\\
9 & 0 & 0 & 0 & 0 & 0 & 0 & -1 & -10 & -61 & -294
\end{array}
\]
\par\end{center}

\begin{center}
Table 5.4: $\mathrm{d}(n,p,4)\propto d^{8}\mathrm{g}(\phi,n,p)/d\phi^{8}|_{\phi=0}$
\[
\begin{array}{rrrrrrrrrrr}
 & p\quad0 & 1 & 2 & 3 & 4 & 5 & 6 & 7 & 8 & 9\\
n\\
0 & 0 & 0 & 0 & 0 & 0 & 0 & 0 & 0 & 0 & 0\\
1 & 14 & 14 & 20 & 35 & 70 & 154 & 364 & 910 & 2380 & 6460\\
2 & -14 & -8 & -5 & 0 & 14 & 56 & 182 & 560 & 1700 & 5168\\
3 & 6 & -3 & -10 & -21 & -42 & -84 & -168 & -330 & -612 & -969\\
4 & -1 & 4 & 4 & 0 & -14 & -56 & -176 & -512 & -1445 & -4028\\
5 & 0 & -1 & 2 & 7 & 14 & 20 & 8 & -91 & -526 & -2147\\
6 & 0 & 0 & -1 & 0 & 6 & 24 & 69 & 168 & 350 & 552\\
7 & 0 & 0 & 0 & -1 & -2 & 1 & 22 & 105 & 378 & 1190\\
8 & 0 & 0 & 0 & 0 & -1 & -4 & -8 & 0 & 84 & 496\\
9 & 0 & 0 & 0 & 0 & 0 & -1 & -6 & -21 & -50 & -54
\end{array}
\]
\par\end{center}

The values of $\mathrm{d}(n,p,k)$ in Tables 5.2, 5.3 and 5.4 can
be generated column by column by this recurrence relation:

\begin{equation}
\begin{array}{l}
\mathrm{d}(0,p,k)=0\\
\mathrm{d}(n,0,k)=\pm\mathrm{d}(n,k-1,1)\\
\mathrm{d}(n,p,k)=\mathrm{d}(n-1,p-1,k)+2\:\mathrm{d}(n,p-1,k)+\mathrm{d}(n+1,p-1,k)
\end{array}
\end{equation}
\newpage{}

The recurrences in Tables 5.2, 5.3 and 5.4 for $k=2$, 3, and 4 are
seeded with the values in the first column ($p=0$) and these values
are given by the values in the $p=k-1$ column in Table 5.1 for $k=1$
but with alternating signs. Because the column entries of Table 5.1
seed the first column of subsequent tables it is convenient to apply
the following recurrence relation which generates the column entries
of Table 5.1 for a particular value of \emph{p:
\begin{equation}
\begin{array}{l}
\mathrm{d}(p+1,p,1)=-1\\
\mathrm{d}(n,p,1)=\mathrm{d}(n+1,p,1)\,\dfrac{n}{(n+1)}\,\dfrac{(p+n+2)}{(p-n+1)}\quad(\mathrm{for}\:n=p,\,p-1,\,...,\,0)
\end{array}
\end{equation}
}Numerical evidence supports the conjecture that the recurrence relation
of Equation 5.6 applies generally for $k>4$. The most important property
of the recurrence relation is that it propagates the lower triangle
of zeroes in each table with a downward shift of one unit for each
unit increase in \emph{k}. This implies that all derivatives of the
Fourier transform evaluated at zero up to the $\mathit{\mathrm{2}k}$th
derivative are zero for $p\leq n-k-1$. In other words the recurrence
relation supports the conjecture that the number of zero-valued even
derivatives, \emph{z}, is given by:

\begin{equation}
z=n-p-1\quad(\mathrm{for}\:p<n)
\end{equation}

The significance of this result cannot be overstated. The validity
of the assertion that the Grace function is well suited for the design
of FIR filters with exceptional flatness in the pass band depends
on the validity of Equation 5.8.\smallskip{}

The Fourier transform of $\mathrm{G}(x,n,p)$, $\mathrm{g}(\phi,n,p)$,
is an even function with $\mathrm{g}(0,n,p)=1$ and its first $n-p-1$
even derivatives at zero are zero therefore its Taylor series is given
by:

\begin{equation}
\mathrm{g}(\phi,n,p)=1+\stackrel[k=n-p]{\infty}{\sum}\dfrac{1}{(2k)!}\left.\frac{d^{2k}\mathrm{g}(\phi,n,p)}{d\phi^{2k}}\right|_{\phi=0}\,\phi^{2k}
\end{equation}

Substituting Equations 4.6, 5.1 and 5.2 into Equation 5.9 gives:

\begin{equation}
\mathrm{g}(\phi,n,p)=1+\stackrel[k=n-p]{\infty}{\sum}\frac{\left[\,\frac{\pi}{2}\,\mathrm{b}(n,p)\,\right]^{2k}}{(2k)!\,4^{p}}\:\mathrm{d}(n,p,k)\,\phi^{2k}
\end{equation}

The factorial term in the denominator of Equation 5.10 makes the summation
convergent for $0\leq\phi\leq1$ which, as noted in Section 3, includes
the pass band. However for moderately large values of \emph{n} the
coefficients of $\phi^{2k}$ tend to blow up rather spectacularly
with increasing values of \emph{k} before they converge to zero. This
results in a severe loss of accuracy due to subtractive cancellation,
a problem that can be addressed by using arbitrary precision arithmetic.\smallskip{}

Because the summation of Equation 5.10 starts with the index $n-p$
the term $\mathrm{d}(n,p,k)$ can be written in the form $\mathrm{d}(n,p,n-p+i)$
where \emph{i} starts at zero. These terms are given by a \emph{i}th
degree polynomial in \emph{n} and \emph{p}. In particular, the equations
for values of \emph{i} of 0, 1 and 2 are:

\begin{equation}
\mathrm{d}(n,p,n-p)=-1
\end{equation}

\begin{equation}
\mathrm{d}(n,p,n-p+1)=2n-4p
\end{equation}

\begin{equation}
\mathrm{d}(n,p,n-p+2)=-2n^{2}+8np-8p^{2}-3n+8p
\end{equation}

In the next section the case where $n\rightarrow\infty$ for a fixed
value of \emph{p} will be considered. The polynomial term with the
highest exponent of \emph{n} (i.e., $n^{i}$) is relevant to this
case and it is given by:

\begin{equation}
\mathrm{d}(n,p,n-p+i)=\frac{-(-2n)^{i}}{i!}+O\left(n^{i-1}\right)
\end{equation}

\newpage{}

\section{The Monotonicity of the Grace Function Fourier Transform}

The relative flatness of the Fourier transform of the Grace function
in the pass band obscures the nature of this function. Insight into
the nature of the Fourier transform in the pass band can be obtained
by considering the following transformation of the Fourier transform:

\begin{equation}
\mathrm{f}(\phi,n,p)=\left[\,1-\mathrm{g}(\phi,n,p)\,\right]^{\tfrac{1}{2(n-p)}}
\end{equation}

Note that $1-\mathrm{g}(\phi,n,p)$ can be calculated using Equation
5.10 without explicitly calculating $\mathrm{g}(\phi,n,p)$ thereby
reducing roundoff error due to  subtractive cancellation. $\mathrm{f_{g}}(\phi,n,p)$
is an odd function so its Taylor series is given by:

\begin{equation}
\mathrm{f}(\phi,n,p)=\stackrel[i=1]{\infty}{\sum}r_{i}(n,p)\,\phi^{2i-1}
\end{equation}

The coefficients, $r_{i}(n,p)$, of this equation can be derived from
the coefficients of the Taylor series of $\mathrm{g}(\phi,n,p)$ given
by Equation 5.10. In particular, the coefficients $r_{1}(n,p)$ and
$r_{2}(n,p)$ are given by:

\begin{equation}
r_{1}(n,p)=\dfrac{\frac{\pi}{2}\,\mathrm{b}(n,p)}{\left[(2(n-p))!\,4^{p}\right]{}^{\tfrac{1}{2(n-p)}}}
\end{equation}

\begin{equation}
r_{2}(n,p)=\dfrac{-r_{1}(n,p)\,\left[\frac{\pi}{2}\,\mathrm{b}(n,p)\right]^{2}(n-2p)}{2\,(n-p)\,(n-p+1)\,(2n-2p+1)}
\end{equation}

The limit of $r_{i}(n,p)$ as $n\rightarrow\infty$ for a fixed value
of \emph{p} will be denoted by $r_{i}^{\infty}$. The first two values
are:

\begin{equation}
\underset{n\rightarrow\infty}{\lim}r_{1}(n,p)=r_{1}^{\infty}=\dfrac{e}{2}
\end{equation}

\begin{equation}
\underset{n\rightarrow\infty}{\lim}r_{2}(n,p)=r_{2}^{\infty}=-\dfrac{e}{8}
\end{equation}

The limit of $\mathrm{f}(\phi,n,p)$ as $n\rightarrow\infty$ for
a fixed value of \emph{p}, denoted by $\mathrm{f^{\infty}}(\phi)$,
is given by:

\begin{equation}
\mathrm{f^{\infty}}(\phi)=\underset{n\rightarrow\infty}{\lim}\mathrm{f}(\phi,n,p)=\stackrel[i=1]{\infty}{\sum}r_{i}^{\infty}\phi^{2i-1}
\end{equation}

The first eight terms of the summation of Equation 6.7 are:

\begin{equation}
\mathrm{f^{\infty}}(\phi)=e\left[\dfrac{1}{2}\,\phi-\dfrac{1}{8}\,\phi^{3}-0\,\phi^{5}-\dfrac{1}{384}\,\phi^{7}-\dfrac{1}{768}\,\phi^{9}-\dfrac{1}{1280}\,\phi^{11}-\dfrac{47}{92160}\,\phi^{13}-\dfrac{731}{2064384}\,\phi^{15}-\cdots\right]
\end{equation}
\newpage{}

The coefficients, $r_{i}^{\infty}$, of Equation 6.8 can be generated
from the following two recurrence relations:

\begin{equation}
\begin{array}{l}
rn=\begin{bmatrix}\begin{array}{r}
-4\\
4\\
0\\
-2
\end{array}\end{bmatrix};\quad rd=\begin{bmatrix}\begin{array}{r}
-8\\
-32\\
1\\
768
\end{array}\end{bmatrix}\quad(\mathrm{for}\:i=1,...\,4)\\
\\
rn_{i}=\dfrac{4\,(i-1)(i-3)(i-4)\,rn_{i-1}-(i-6)\,rn_{i-3}}{i\,(2i-7)}\quad(\mathrm{for}\:i=5,...\,\infty)\\
\\
rd_{i}=\dfrac{2\,(i-1)(i-4)\,rd_{i-1}}{(i-3)}\quad(\mathrm{for}\:i=5,...\,\infty)\\
\\
r_{i}^{\infty}=e\:\dfrac{rn_{i}}{rd_{i}}\quad(\mathrm{for}\:i=1,...\,\infty)
\end{array}
\end{equation}

The form of the recurrence relations make $rn_{i}$ and $rd_{i}$
holonomic (polynomially recursive) sequences. The values of $rn_{i}$
and $rd_{i}$ increase exponentially with increasing $i$ which will
cause an overflow with 64-bit arithmetic for $i$ greater than 150.
This problem is solved by scaling the last three values of $rn_{i}$
and $rd_{i}$ when the value of $rd_{i}$ approaches the overflow
value.

\smallskip{}

The series in Equation 6.8 converges for $0\leq\phi\leq1$. Therefore
the radius of convergence of the series coincides with the range of
values of $\phi$ which define the pass band. The rate of convergence
of the series is very slow for $\phi=1$. The partial sums of the
series for $\mathrm{f_{g}^{\infty}}(1)$ and $\mathrm{\mathit{d}f_{g}^{\infty}}(\phi)/d\phi|_{\phi=1}$
are given by the following series:

\begin{equation}
\stackrel[i=1]{j}{\sum}r_{i}^{\infty}=1+\dfrac{1}{\sqrt{\pi j}}\left[\dfrac{1}{6}j^{-1}+\dfrac{3}{80}j^{-2}-\dfrac{65}{5376}j^{-3}-\dfrac{2681}{55296}j^{-4}-\dfrac{55453}{720896}j^{-5}+\cdots\right]
\end{equation}

and

\begin{equation}
\stackrel[i=1]{j}{\sum}r_{i}^{\infty}(2i-1)=\dfrac{1}{\sqrt{\pi j}}\left[1+\dfrac{1}{8}j^{-1}+\dfrac{1}{128}j^{-2}-\dfrac{355}{3072}j^{-3}-\dfrac{19775}{98304}j^{-4}-\dfrac{326069}{1310720}j^{-5}+\cdots\right]
\end{equation}

Taking the limit as $j\rightarrow\infty$ of these partial sums of
the series for $\mathrm{f_{g}^{\infty}}(1)$ and $\mathrm{\mathit{d}f_{g}^{\infty}}(\phi)/d\phi|_{\phi=1}$
gives:

\begin{equation}
\mathrm{f^{\infty}}(1)=1
\end{equation}

and

\begin{equation}
\mathrm{\mathit{d}f^{\infty}}(\phi)/d\phi|_{\phi=1}=0
\end{equation}

The elegant simplicity of these two results is notable given that
elegant simplicity is not an attribute of the recurrence relations
of Equation 6.9.\smallskip{}

In the vicinity of $\phi=1$ the following series, with a change of
variable, has a much faster rate of convergence than the series of
equation 6.8:

\begin{equation}
\theta\equiv\sqrt{2\left(1-\phi\right)}
\end{equation}

\begin{equation}
\mathrm{f^{\infty}}(\theta)=1-\frac{1}{3}\theta^{3}-\frac{3}{40}\theta^{5}+\frac{8}{144}\theta^{6}-\frac{23}{896}\theta^{7}+\frac{192}{7680}\theta^{8}-\frac{1331}{82944}\theta^{9}+\frac{12224}{1075200}\theta^{10}-\frac{132879}{16220160}\theta^{11}\cdots
\end{equation}

The denominators of the coefficients in this series are a holonomic
sequence. However an analysis of the first 36 coefficients of this
series suggests that the numerators only asymptotically approach a
holonomic sequence. Therefore a recurrence relation for the numerators
remains elusive.

\newpage{}

$\mathrm{f^{\infty}}(\phi)$ is plotted in Figures 6.1 and 6.2 in
black along with plots of $\mathrm{f}(\phi,n,p)$ for selected values
of \emph{n} and \emph{p} where the corresponding ratios of the values
of \emph{p} to \emph{n} are the same in the two figures.
\begin{center}
Figure 6.1: Plot of $\mathrm{f^{\infty}}(\phi)$ and $\mathrm{f}(\phi,n,p)$
for $n=10$ and $p=0,5,6,7,8,9$
\par\end{center}

\begin{center}
\includegraphics[scale=0.6]{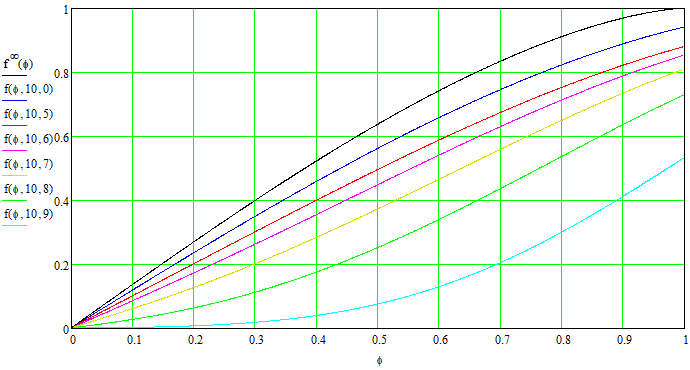}
\par\end{center}

\begin{center}
Figure 6.2: Plot of $\mathrm{f^{\infty}}(\phi)$ and $\mathrm{f}(\phi,n,p)$
for $n=100$ and $p=0,50,60,70,80,90$
\par\end{center}

\begin{center}
\includegraphics[scale=0.6]{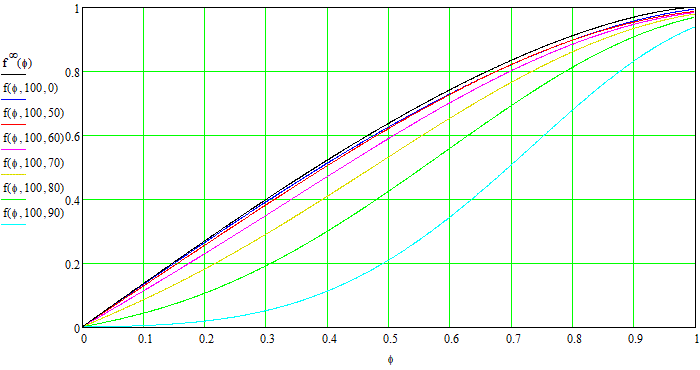}
\par\end{center}

These figures can be put in better context by inverting Equation 6.1
to give the Fourier transform, $\mathrm{g}(\phi,n,p)$, in terms of
$\mathrm{f}(\phi,n,p)$:

\begin{equation}
\mathrm{g}(\phi,n,p)=1-\mathrm{f}(\phi,n,p)^{2(n-p)}\quad(\mathrm{for}\:p<n)
\end{equation}

It follows from this equation that $\mathrm{g}(\phi,n,p)$ decreases
monotonically in the range $0\leq\phi\leq1$ iff $\mathrm{f}(\phi,n,p)$
increases monotonically. Figures 6.1 and 6.2 illustrate the monotonic
behaviour of the function $\mathrm{f}(\phi,n,p)$. It is shown in
Appendix 3 that in the vicinity of $\phi=0$, $\mathrm{f}(\phi,n,p)$
is concave for $p\leq\nicefrac{n}{2}$ and convex for $p>\nicefrac{n}{2}$
for all values of \emph{n}.\newpage{}

A comparison of Figures 6.1 and 6.2 might suggest that $\mathrm{f^{\infty}}(\phi)$
is an upper bound on $\mathrm{f}(\phi,n,p)$. Figure 6.3 for $n=1000$
shows that this is not the case. Values of $\mathrm{f}(\phi,1000,500)$
exceed the values of $\mathrm{f^{\infty}}(\phi)$ for $0.258<\phi<0.972$.
An analysis of this anomaly is presented in Appendix 3.
\begin{center}
Figure 6.3: Plot of $\mathrm{f^{\infty}}(\phi)$ and $\mathrm{f}(\phi,n,p)$
for $n=1000$ and $p=0,500,600,700,800,900$
\par\end{center}

\begin{center}
\includegraphics[scale=0.6]{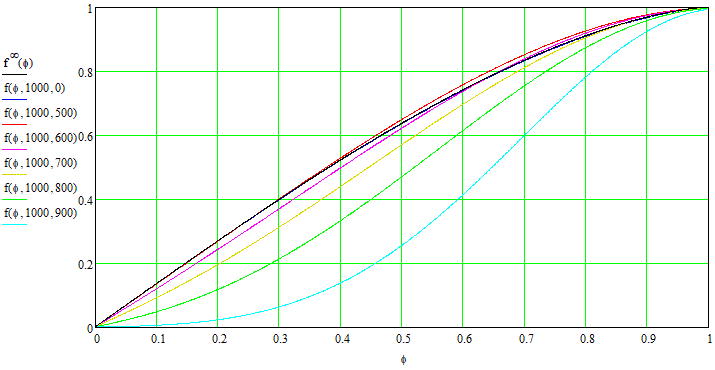}
\par\end{center}

All of the curves in Figures 6.1, 6.2 and 6.3 increase monotonically
for $0\leq\phi\leq1$ and they tend to converge with increasing \emph{n}.
This consistent behaviour of the function $\mathrm{f}(\phi,n,p)$
strongly supports the conjecture that ripple-free behaviour is a property
of the Fourier transform of the Grace function in the pass band.\smallskip{}

Equation 5.8 shows that the local behaviour of the Fourier transform
in the neighbourhood of $\phi=0$ is largely influenced by the value
of $n-p$. Equation 6.16 shows that the global behaviour of the Fourier
transform in the entire pass band is also largely influenced by the
value of $n-p$. Because $\mathrm{f^{\infty}}(\phi)<1$ over the range
$0\leq\phi<1$ it follows from Equation 6.16 that:

\begin{equation}
\mathrm{g}^{\infty}(\phi)=\underset{n-p\rightarrow\infty}{\lim}\mathrm{g}(\phi,n,p)=1\quad(\mathrm{for}\:0\leq\phi<1)
\end{equation}

However, the result that $\mathrm{f^{\infty}}(1)=1$ implies that
$\mathrm{g}^{\infty}(1)=0$. Therefore $\mathrm{g}^{\infty}(\phi)$
is a step function which is the Fourier transform of the sinc function.
This, in turn, implies that the Grace function approaches a scaled
sinc function as \emph{$n-p\rightarrow\infty$}.\newpage{}

\section{The Grace Filter and Its Properties}

The discrete Grace filter is derived from the Grace function, Equation
3.1, defined for \emph{x} on the interval $\left[-1,1\right]$. The
$2m+1$ coefficients of the Grace filter are given by:

\begin{equation}
c_{i}=\frac{\mathrm{G}(x_{i},n,p)}{\stackrel[i=-m]{m}{\sum}\mathrm{G}(x_{i},n,p)}
\end{equation}

where:

\begin{equation}
x_{i}=\dfrac{i}{m}\quad(\mathrm{for}\:i=-m,...\,m)
\end{equation}

The Grace filter is symmetric with an odd number of coefficients making
it a type I linear-phase FIR filter. The frequency response of the
Grace filter approaches the Fourier transform, $\mathrm{g}(\phi,n,p)$,
of the Grace function as $m\rightarrow\infty$. Two features of the
frequency response, the transition band rolloff (in dB/octave) and
the amplitude of the first stop band sidelobe, tend to rapidly converge
to the values of the Fourier transform with increasing values of \emph{m}.
Tables 7.1 and 7.2 give the limiting values of the transition band
rolloff and the amplitude of the first stop band sidelobe for a wide
range of values of the filter parameters \emph{n} and \emph{p}. Because
\emph{p} is constrained to the range $0\leq p<n$ the values of \emph{p}
are expressed as a fraction of the values of \emph{n.}
\begin{center}
Table 7.1: The limiting transition band rolloff in dB/octave for the
Grace filter as $m\rightarrow\infty$
\[
\begin{array}{rrrrrrrrrrrr}
 & \nicefrac{p}{n}\quad0.0 & 0.1 & 0.2 & 0.3 & 0.4 & 0.5 & 0.6 & 0.7 & 0.8 & 0.9 & 1.0\\
n\:\\
10 & 25.4 & 20.4 & 18.2 & 16.5 & 15.2 & 14.1 & 13.1 & 12.3 & 11.6 & 11.0 & 10.5\\
20 & 41.5 & 31.6 & 27.9 & 25.2 & 23.0 & 21.2 & 19.8 & 18.5 & 17.5 & 16.5 & 15.7\\
30 & 55.1 & 40.7 & 35.7 & 32.1 & 29.2 & 26.9 & 24.9 & 23.3 & 22.0 & 20.8 & 19.8\\
50 & 78.8 & 55.9 & 48.5 & 43.2 & 39.1 & 35.8 & 33.2 & 31.0 & 29.2 & 27.6 & 26.2\\
70 & 99.5 & 68.7 & 59.2 & 52.4 & 47.3 & 43.2 & 39.9 & 37.3 & 35.0 & 33.1 & 31.5\\
100 & 127.4 & 85.5 & 73.0 & 64.2 & 57.6 & 52.5 & 48.5 & 45.2 & 42.4 & 40.1 & 38.1\\
150 & 168.6 & 109.4 & 92.3 & 80.5 & 71.9 & 65.3 & 60.2 & 56.1 & 52.6 & 49.7 & 47.2\\
200 & 205.5 & 130.2 & 108.8 & 94.4 & 84.0 & 76.2 & 70.1 & 65.2 & 61.2 & 57.8 & 54.9\\
300 & 271.4 & 166.0 & 136.8 & 117.7 & 104.3 & 94.4 & 86.8 & 80.6 & 75.6 & 71.4 & 67.8\\
500 & 384.8 & 224.5 & 181.7 & 154.8 & 136.6 & 123.3 & 113.2 & 105.1 & 98.5 & 93.0 & 88.3\\
700 & 483.9 & 273.3 & 218.4 & 185.1 & 162.8 & 146.8 & 134.6 & 125.0 & 117.1 & 110.5 & 104.9\\
1000 & \:616.8 & \:335.9 & \:264.9 & \:223.3 & \:195.9 & \:176.4 & \:161.7 & \:150.0 & \:140.5 & \:132.6 & \:125.9
\end{array}
\]
\par\end{center}

\begin{center}
Table 7.2: The limiting amplitude of the first stop band sidelobe
in dB as $m\rightarrow\infty$
\[
\begin{array}{rrrrrrrrrrrr}
 & \nicefrac{p}{n}\quad0.0 & 0.1 & 0.2 & 0.3 & 0.4 & 0.5 & 0.6 & 0.7 & 0.8 & 0.9 & 1.0\\
n\:\\
10 & -12.7 & -17.4 & -22.1 & -26.6 & -31.0 & -35.3 & -39.5 & -43.5 & -47.5 & -51.4 & -55.3\\
20 & -12.1 & -18.2 & -24.4 & -30.7 & -37.1 & -43.4 & -49.7 & -55.9 & -62.2 & -68.5 & -74.7\\
30 & -11.9 & -18.9 & -26.3 & -34.0 & -41.8 & -49.8 & -57.9 & -66.1 & -74.5 & -82.8 & -91.3\\
50 & -11.7 & -20.2 & -29.3 & -39.1 & -49.6 & -60.5 & -71.9 & -83.7 & -95.8 & -108.2 & -120.7\\
70 & -11.6 & -21.2 & -31.7 & -43.4 & -56.2 & -69.8 & -84.3 & -99.4 & -115.1 & -131.2 & -147.8\\
100 & -11.5 & -22.4 & -34.8 & -49.0 & -64.8 & -82.3 & -101.0 & -121.0 & -141.9 & -163.6 & -186.0\\
150 & -11.5 & -24.1 & -39.1 & -56.9 & -77.6 & -100.9 & -126.6 & -154.2 & -183.6 & -214.3 & -246.3\\
200 & -11.4 & -25.5 & -42.7 & -63.8 & -89.1 & -118.1 & -150.4 & -185.6 & -223.2 & -262.9 & -304.2\\
300 & -11.4 & -27.8 & -48.9 & -76.2 & -110.0 & -149.9 & -195.3 & -245.3 & -299.2 & -356.4 & -416.4\\
500 & -11.3 & -31.3 & -59.1 & -97.7 & -147.8 & -208.8 & -279.5 & -358.4 & -444.4 & -536.3 & -633.2\\
700 & -11.3 & -34.2 & -67.9 & -117.0 & -182.8 & -264.4 & -360.0 & -467.6 & -585.3 & -711.6 & -845.1\\
1000 & -11.3 & -37.8 & -79.7 & -143.9 & -232.6 & -344.6 & -477.2 & -627.4 & -792.5 & -970.1 & -1158.3
\end{array}
\]
\newpage{}
\par\end{center}

A comparison of Tables 7.1 and 7.2 shows the trade-off in selecting
$\mathit{p}$ for a given value of $\mathit{n}$. Small values of
$\mathit{p}$ give more favourable transition band values while large
values of $\mathit{p}$ give more favourable stop band values. But
because large values of $\mathit{n}$ are favourable to both the transition
band rolloff and the first sidelobe amplitude it will generally be
possible to select values of $\mathit{n}$ and $\mathit{p}$ that
meet specifications on both the transition band rolloff and the first
sidelobe amplitude.\smallskip{}

There are two common conventions for expressing frequency in the context
of FIR filter design. One convention associates $\nicefrac{1}{2}$
with the Nyquist frequency while the other convention, used in this
study, associates 1 with the Nyquist frequency. For a type I FIR filter
the frequency response, h, as a function of frequency, \emph{f}, scaled
on the interval $\left[0,1\right]$ is given by:

\begin{equation}
\mathrm{h}(f)=\stackrel[i=-m]{m}{\sum}c_{i}\,\cos(i\,\pi\,f)=\stackrel[i=-m]{m}{\sum}c_{i}\,\cos(x_{i}\,m\,\pi\,f)
\end{equation}

The derivative of the frequency response with respect to frequency
is given by:

\begin{equation}
\dfrac{d\mathrm{h}(f)}{df}=-m\pi\stackrel[i=-m]{m}{\sum}c_{i}\,x_{i}\sin(x_{i}\,m\,\pi\,f)
\end{equation}

The frequency response is an even function in \emph{f} so the odd
derivatives of the frequency response at zero frequency are all zero.
The even derivatives at zero frequency are given by:

\begin{equation}
\left.\frac{d^{2k}\mathrm{h(\mathit{f})}}{df^{2k}}\right|_{f=0}=\left(-1\right)^{k}(m\pi)^{2k}\stackrel[i=-m]{m}{\sum}c_{i}\,x_{i}^{2k}\quad(\mathrm{for}\:k=1,\,2,\,3,...)
\end{equation}

The term $(m\pi)^{2k}$ can be extremely large therefore it will be
convenient to work with the following product:

\begin{equation}
(m\pi)^{-2k}\left.\frac{d^{2k}\mathrm{h(\mathit{f})}}{df^{2k}}\right|_{f=0}=\left(-1\right)^{k}\stackrel[i=-m]{m}{\sum}c_{i}\,x_{i}^{2k}\quad(\mathrm{for}\:k=1,\,2,\,3,...)
\end{equation}

The integral of the squared frequency response over the frequency
range will be called the reference frequency, $f_{r}$, given by:

\begin{equation}
f_{r}=\int_{0}^{1}\mathrm{h}\left(f\right)^{2}df=\stackrel[i=-m]{m}{\sum}c_{i}^{2}
\end{equation}

If the squared frequency response is reasonably anti-symmetric around
the coordinates $(f_{c},\nicefrac{1}{2})$ where $f_{c}$ is the cutoff
frequency, then the reference frequency will be close to the cutoff
frequency. The significance of this is that, for the Grace filter,
the following equation gives the limiting relationship between the
values of \emph{n}, \emph{p}, \emph{m} and $f_{r}$

\begin{equation}
\underset{m\rightarrow\infty}{\lim}m\,f_{r}=\mathrm{b}(n,p)\quad(\mathrm{for}\:p<n)
\end{equation}

where $\mathrm{b}(n,p)$ is given by Equation 3.6. $m\,f_{r}$ tends
to rapidly converge to the limiting value with an increasing value
of \emph{m}. This gives the following good approximation for the product
of \emph{m} and the cutoff frequency, $f_{c}$, where $\mathrm{b}(n,p)$
can be evaluated using the approximation of Equation 4.8:

\begin{equation}
m\,f_{c}\approx\mathrm{b}(n,p)
\end{equation}

A strategy for the design of a Grace filter is to use Tables 7.1 and
7.2 to determine the values of the filter parameters \emph{n} and
\emph{p} that will meet given specifications on both the transition
band rolloff and the amplitude of the first stop band sidelobe. With
these values of \emph{n} and \emph{p} Equation 7.9 can then used to
determine the value of the filter parameter \emph{m} that will meet
a given specification on the filter cutoff frequency. The Grace filter
is implemented in the program Grace in Appendix 1.\newpage{}

A comparison of Equation 3.2 for the Fourier transform of the Grace
function with Equation 7.3 for the frequency response of the Grace
filter suggests that the summation of Equation 7.3 can be thought
of as numerical integration of the integral of Equation 3.2 with a
stepsize of $\nicefrac{1}{m}$. In Section 5 it was shown that the
properties of the Fourier transform led to the important result that
the first $n-p-1$ even derivatives of the Fourier transform evaluated
at zero are zero. However it cannot be expected that the approximate
numerical integration of the integral will preserve this important
result. The values of the first $\mathit{z=n-p-\mathrm{1}}$ even
derivatives of the frequency response of the Grace filter at zero
frequency are given by this series in \emph{m}:

\begin{equation}
(m\pi)^{-2k}\left.\frac{d^{2k}\mathrm{h(\mathit{f})}}{df^{2k}}\right|_{f=0}=\frac{\alpha(n,p)}{m^{p+\nicefrac{3}{2}}}\left[1-\frac{\beta(n,p,k)}{m}-\frac{\gamma(n,p,k)}{m^{2}}+\cdots\right]\quad(\mathrm{for}\:k=1,2,...\,z)
\end{equation}

where:

\begin{equation}
\alpha(n,p)=(-1)^{n+[p-\mathrm{mod}(p,2)]/2}\,\frac{\sqrt{8}}{\pi}\,n\stackrel[i=1]{\infty}{\sum}\frac{1}{i^{p+\nicefrac{3}{2}}}\stackrel[j=0]{p}{\prod}\frac{j+\nicefrac{1}{2}}{\pi}
\end{equation}

\begin{equation}
\beta(n,p,k)=(-1)^{p}\,\frac{\left(p+\nicefrac{3}{2}\right)}{12\pi}\stackrel[i=1]{\infty}{\sum}\frac{J1_{i}}{i^{p+\nicefrac{5}{2}}}\left[8n^{2}+3p+12\left(k-1\right)+6-\frac{1}{2}\right]
\end{equation}

\begin{align}
\gamma(n,p,k) & =\frac{\left(p+\nicefrac{3}{2}\right)\left(p+\nicefrac{5}{2}\right)}{480\pi^{2}}\stackrel[i=1]{\infty}{\sum}\frac{J2_{i}}{i^{p+\nicefrac{7}{2}}}\;\times\nonumber \\
 & \left[8n^{2}\left(8n^{2}+10p+40k-25\right)+5p\left(3p+24k-16\right)+20\left(k-1\right)\left(12k-7\right)-8+\frac{1}{4}\right]
\end{align}

The dominant term in Equation 7.10 is $m^{-(p+\nicefrac{3}{2)}}$.
For moderately large values of \emph{m} and \emph{p} the values of
the first $\mathit{z=n-p-\mathrm{1}}$ even derivatives of the frequency
response of the Grace filter will be negligible. In those cases where
the values of these derivatives are not negligible the algorithm of
the next section will determine the smallest adjustments to the filter
coefficients necessary to make the values of the derivatives exactly
zero.\smallskip{}
The summation terms in the equations for $\alpha,\,\beta\,\mathrm{and}\,\gamma$
are Dirichlet series usually associated with analytic number theory.
The appearance of these functions in this context is something of
a mystery. The summation term in the equation for $\alpha$ is the
Riemann zeta function with argument $p+\nicefrac{3}{2}$. The coefficients
$J1_{i}$ and $J2_{i}$ in the summation terms in the equations for
$\beta$ and $\gamma$ are the Dirichlet inverses of the Jordan totient
functions, $J_{1}$and $J_{2}$. These coefficients are listed as
sequences A023900 and A046970 in the On-line Encyclopedia of Integer
Sequences (OEIS) {[}3{]}.

\smallskip{}

As an aside, the Grace filter coefficients, \emph{c}, (Equation 7.1)
will be considered for given values of \emph{m}, \emph{n} and \emph{p.}
The elements of the matrix \emph{A} are defined as:

\begin{equation}
A_{i,j}=x_{i}^{2(j-1)}\quad(\mathrm{for}\:i=-m,...\,m\:\mathrm{and}\:j=1,...\,n)
\end{equation}

The elements of the diagonal matrix \emph{W} are defined as:

\begin{equation}
W_{i,i}=\left(1-x_{i}^{2}\right)^{p+\nicefrac{1}{2}}
\end{equation}

With these definitions the following empirical result implies that
the Grace filter coefficient vector \emph{c} is an eigenvector of
the matrix $WA(A^{T}WA)^{-1}A^{T}$:

\begin{equation}
WA\left(A^{T}WA\right)^{-1}A^{T}c=c
\end{equation}

The matrix \emph{A} appears in the least-squares estimation of the
parameters of an even polynomial over the ranges in \emph{x} and \emph{j}
of Equation 7.14. In this case the coefficients of the polynomial
also satisfy Equation 7.16. This suggests an interesting connection
between the Grace filter and filters such as Savitzky-Golay which
are based on a polynomial fit to a segment of data.

\newpage{}

\section{The Discretization Compensation Algorithm}

The first $n-p-1$ even derivatives of the Fourier transform of the
Grace function evaluated at zero are zero. Unfortunately the frequency
response of the Grace filter does not share this important property
because of the filter's discrete nature. The dominant term in Equation
7.10 for the values of these derivatives is $m^{-(p+\nicefrac{3}{2)}}$.
For moderately large values of \emph{m} and \emph{p} the values of
the first $\mathit{z=n-p-\mathrm{1}}$ even derivatives of the frequency
response of the Grace filter at zero frequency will be negligible.
In those cases where the values of these derivatives are not negligible
the algorithm of this section will determine the smallest adjustments
to the filter coefficients necessary to make the values of the derivatives
exactly zero.\smallskip{}

In Equations 7.5 and 7.6 the filter coefficients, $c_{i}$, are considered
as constants. If they are considered as variables then the partial
derivative of $(m\pi)^{-2k}d^{2k}\mathrm{h}(f,c)/df^{2k}|_{f=0}$
with respect to $\mathit{c_{i}}$ is given by:

\begin{equation}
(m\pi)^{-2k}\left.\frac{\partial}{\partial c_{i}}\frac{d^{2k}\mathrm{h(\mathit{f,c})}}{df^{2k}}\right|_{f=0}=\left(-1\right)^{k}x_{i}^{2k}\quad(\mathrm{for}\:k=1,\,2,\,3,...)
\end{equation}

Because the relationship between the filter coefficients and the even
derivatives of the frequency response at zero frequency is linear
the adjustments to $\mathit{c}$ required to make the first $\mathit{z}=n-p-1$
even derivatives of $\mathrm{h}(f,c)$ exactly zero at zero frequency
can be determined by solving a set of linear equations. In principle
the number of coefficients to be adjusted can be as low as the value
of \emph{z}. However, there is merit in making the adjustments as
small as possible which means that all of the coefficients in $\mathit{c}$
will be adjusted. The set of linear equations for the adjustments
in $\mathit{c}$, $\Delta c$, can be written as:

\begin{equation}
A\Delta c=b
\end{equation}

The filter coefficients, $\mathit{c}$, sum to one so the adjustments
to these coefficients, $\Delta c$, must sum to zero. This constraint
is imposed in the first row of matrix \emph{A} and vector \emph{b}:

\begin{equation}
A_{1,i}=1\quad(\mathrm{for}\:i=-m,...\,m)
\end{equation}

\begin{equation}
b_{1}=0
\end{equation}

$b_{1+k}$ is the desired change in the value of the derivative, $(m\pi)^{-2k}d^{2k}\mathrm{h}(f,c)/df^{2k}|_{f=0}$.
The objective is to make the value of the derivative zero so, from
Equation 7.6:

\begin{equation}
b_{1+k}=-(m\pi)^{-2k}\left.\frac{d^{2k}\mathrm{h}(f,c)}{df^{2k}}\right|_{f=0}=-\left(-1\right)^{k}\stackrel[i=-m]{m}{\sum}c_{i}\,x_{i}^{2k}\quad(\mathrm{for}\:k=1,...\,z)
\end{equation}

Equations 8.4 and 8.5 for the vector \emph{b} can be written in matrix
form as:

\begin{equation}
b=B\,c
\end{equation}

where:

\begin{equation}
B_{j,i}=\left\{ \begin{array}{cc}
0 & \mathrm{for}\,j=1\\
-(-1)^{j-1}x_{i}^{2(j-1)} & \mathrm{for}\,j>1
\end{array}\right.\quad(\mathrm{for}\:i=-m,\,...\,m\:\mathrm{and}\:j=1,...\,z+1)
\end{equation}

The elements of the row of $\mathit{A}$, $A_{1+k,i}$, corresponding
to $b_{1+k}$ are the partial derivatives of the scalar $(m\pi)^{-2k}d^{2k}\mathrm{h}(f,c)/df^{2k}|_{f=0}$
with respect to the elements of the vector $\mathit{c}$ and these
partial derivatives are given by Equation 8.1. Therefore the elements
of $\mathit{A}$, including the first row, are given by:

\begin{equation}
A_{j,i}=\left(-1\right)^{j-1}x_{i}^{2(j-1)}\quad(\mathrm{for}\:i=-m,\,...\,m\:\mathrm{and}\:j=1,...\,z+1)
\end{equation}

The matrix \emph{A} has a very regular structure which will be exploited
later.\pagebreak{}

Combining Equations 8.2 and 8.6 gives:

\begin{equation}
A\Delta c=B\,c
\end{equation}

The common factor of $(-1)^{j-1}$ in the rows of \emph{A} and \emph{B}
will be divided out in Equation 8.9 making all of the elements of
matrix \emph{A} positive.\smallskip{}

This scheme introduces the complication that is that there will generally
be more unknowns than there are equations making the set of equations
under-determined. There is a least-squares solution to an under-determined
set of linear equations. This solution satisfies the set of equations
while minimizing the sum of squares of the elements of the solution
vector which in this case is the vector $\Delta c$:

\begin{equation}
\Delta c=A^{T}\left(AA^{T}\right)^{-1}B\,c
\end{equation}

This least-squares solution satisfies the requirement that the adjustments
to the coefficients, $\Delta c$, be as small as possible. For the
Grace filter the value of $\mathit{c_{\pm m}}$ is zero. This condition
will be imposed on the solution vector $\Delta c$ by adding a weight
matrix, $\mathit{W}$, to Equation 8.10:

\begin{equation}
\Delta c=WA^{T}\left(AWA^{T}\right)^{-1}B\,c
\end{equation}

$\mathit{W}$ is a diagonal matrix of weights whose diagonal elements
are symmetric and zero at the endpoints, i.e., $W_{i,i}=W_{-i,-i}$
and $W_{m,m}=0$. Premultiplying both sides of Equation 8.11 by matrix
\emph{A} gives Equation 8.9 therefore solutions of Equation 8.11 will
satisfy Equation 8.9 for any nonsingular weight matrix \emph{W. }It
will be shown that an appropriate choice for the diagonal elements
of \emph{W} is given by this equation:

\begin{equation}
W_{i,i}=\frac{2}{m\pi}\sqrt{1-x_{i}^{2}}\quad(\mathrm{for}\:i=-m,...\,m)
\end{equation}

With these specific weights the constraint $m>n-p-1$ must be imposed
to avoid singularity in the matrix $AWA^{T}$ of Equation 8.11. Furthermore
the case where $m=n-p$ results in a degenerate solution where only
one coefficient, $(c+\Delta c)_{0}$, is nonzero. The parameter $z=n-p-1$
must be at least one giving the following constraints on the parameters
\emph{m}, \emph{n} and \emph{p}:

\begin{equation}
m>n-p>1
\end{equation}

The matrix $AWA^{T}$ becomes increasingly ill-conditioned as its
rank, $z+1$, increases. The application of matrix preconditioning
is an effective method of addressing the problem of ill-conditioning.
The preconditioning matrix, \emph{P}, is chosen such that:

\begin{equation}
PAWA^{T}P^{T}\approx I
\end{equation}

With this Equation 8.11 can be written as:

\begin{equation}
\Delta c=WA^{T}P^{T}\left(PAWA^{T}P^{T}\right)^{-1}PB\,c
\end{equation}

The validity of this equation can be verified by multiplying both
sides by $P^{-1}PA$ which yields Equation 8.9. To derive a preconditioning
matrix satisfying the approximation of Equation 8.14 the limit of
the elements of $AWA^{T}$ as $m\rightarrow\infty$ will be considered.
The matrix \emph{C} will be defined as:

\begin{equation}
C_{u,v}=\underset{m\rightarrow\infty}{\lim}\left(AWA^{T}\right)_{u,v}=\frac{2}{\pi}\int_{-1}^{1}x^{2\left(u+v-2\right)}\sqrt{1-x^{2}}\,dx\quad(\mathrm{for}\:u,v=1,\,2,...\,z+1)
\end{equation}

The integral can be solved analytically to give:

\begin{equation}
C_{u,v}=-2\stackrel[k=1]{u+v-1}{\prod}\frac{2k-3}{2k}\quad(\mathrm{for}\:u,v=1,\,2,...\,z+1)
\end{equation}
\newpage{}

The limit of Equation 8.16 converges relatively rapidly giving the
following approximation for $AWA^{T}$:

\begin{equation}
AWA^{T}\approx C
\end{equation}

The matrix \emph{C} and its inverse are symmetric so the Cholesky
factorization of \emph{$C^{-1}$ }can be written as:

\begin{equation}
C^{-1}=P^{T}P
\end{equation}

where \emph{P}, the desired preconditioning matrix, is a lower triangular
matrix with integer elements. Table 8.1 gives the first 10 rows of
the preconditioning matrix \emph{P}:
\begin{center}
Table 8.1: The first 10 rows of the preconditioning matrix \emph{P}.
\par\end{center}

\[
\begin{array}{rrrrrrrrrr}
1 & 0 & 0 & 0 & 0 & 0 & 0 & 0 & 0 & 0\\
-1 & 4 & 0 & 0 & 0 & 0 & 0 & 0 & 0 & 0\\
1 & -12 & 16 & 0 & 0 & 0 & 0 & 0 & 0 & 0\\
-1 & 24 & -80 & 64 & 0 & 0 & 0 & 0 & 0 & 0\\
1 & -40 & 240 & -448 & 256 & 0 & 0 & 0 & 0 & 0\\
-1 & 60 & -560 & 1792 & -2304 & 1024 & 0 & 0 & 0 & 0\\
1 & -84 & 1120 & -5376 & 11520 & -11264 & 4096 & 0 & 0 & 0\\
-1 & 112 & -2016 & 13440 & -42240 & 67584 & -53248 & 16384 & 0 & 0\\
1 & -144 & 3360 & -29568 & 126720 & -292864 & 373736 & -245760 & 65536 & 0\\
-1 & 180 & -5280 & 59136 & -329472 & 1025024 & -1863680 & 1966080 & -1114112 & 262144
\end{array}
\]

Unfortunately matrix \emph{P} is useless as preconditioning matrix.
When applied as a preconditioning matrix the alternating signs of
the row elements of \emph{P} will lead to catastrophic subtractive
cancellation as the number of rows increases. At this point a mathematical
rabbit will be pulled from a hat. The elements of each row of matrix
\emph{P} are the coefficients of even-order Chebyshev polynomials
of the second kind. The elements of each column of matrix \emph{A}
are a power series in $x^{2}$. Therefore the elements of the matrix
products \emph{$PA$} and $A^{T}P^{T}$ of Equation 8.15 are given
by the values of Chebyshev polynomials of the second kind:

\begin{equation}
\left(PA\right)_{j,i}=\stackrel[k=1]{z+1}{\sum}P_{j,k}A_{k,i}=U_{2(j-1)}(x_{i})\quad(\mathrm{for}\:i=-m,...\,m\:\mathrm{and}\:j=1,...\,z+1)
\end{equation}

The Chebyshev polynomials of the second kind, $U_{j}(x)$, are efficiently
evaluated using the recurrence relation:

\begin{equation}
\begin{array}{l}
U_{0}(x)=1\\
U_{1}(x)=2x\\
U_{j}(x)=2x\,U_{j-1}(x)-U_{j-2}(x)
\end{array}
\end{equation}

To evaluate Equation 8.15 with as much accuracy as possible the matrix
$PAWA^{T}P^{T}$ is factored as:

\begin{equation}
PAWA^{T}P^{T}=\left(PA\sqrt{W}\right)\left(PA\sqrt{W}\right)^{T}
\end{equation}

where the diagonal elements of $\sqrt{W}$ are the square roots of
the diagonal elements of \emph{W}. This gives:

\begin{equation}
\left(PA\sqrt{W}\right)_{j,i}=U_{2(j-1)}(x_{i})\sqrt{\frac{2}{m\pi}\sqrt{1-x_{i}^{2}}}\quad(\mathrm{for}\:i=-m,...\,m\:\mathrm{and}\:j=1,...\,z+1)
\end{equation}

The preconditioning matrix \emph{P} also multiplies the matrix \emph{B}
whose elements are the negative of the elements of matrix \emph{A}
except for the first row where $B_{1,i}=0$ whereas $A_{1,i}=1$.
This is a minor complication which is resolved in the form of this
equation: 

\begin{equation}
\left(PBc\right)_{j}=\stackrel[i=-m]{m}{\sum}\left[\stackrel[k=1]{z+1}{\sum}P_{j,k}B_{k,i}\right]c_{i}=-(-1)^{j}-\stackrel[i=-m]{m}{\sum}U_{2(j-1)}(x_{i})\,c_{i}\quad(\mathrm{for}\:j=1,...\,z+1)
\end{equation}
\newpage To maintain as much accuracy as possible in the evaluation
of Equation 8.15 the matrix $\left(PA\sqrt{W}\right)^{T}$ is factored
by singular value decomposition (SVD) into the matrices \emph{U},
\emph{S} and \emph{V}:

\begin{equation}
\left(PA\sqrt{W}\right)^{T}=USV^{T}
\end{equation}

The transpose of $PA\sqrt{W}$ is factored because some SVD algorithms
require that the number of matrix rows not be less than the number
of columns. \emph{S} is a diagonal matrix of singular values. With
the preconditioning of matrix \emph{P}, most of the singular values
will be close to one. The notation \emph{U} is standard with SVD,
not to be confused with the notation $U_{j}(x)$ for Chebyshev polynomials
of the second kind. The value of $\Delta c$ is now given by:

\begin{equation}
\Delta c=\sqrt{W}\,US^{-1}V^{T}\left(PBc\right)
\end{equation}

It is computationally efficient to evaluate the right hand side of
this equation from right to left. Also note that it is not necessary
to store any of the matrices \emph{A}, \emph{B}, \emph{P} or \emph{W}.
Only the matrix $(PA\sqrt{W})^{T}$ is stored with elements given
by Equation 8.23. Similarly, the vector $PBc$ is stored with elements
given by Equation 8.24.\smallskip{}

The factor of $\nicefrac{2}{m\pi}$ in the weights given by Equation
8.12 makes the matrix $PAWA^{T}P^{T}$ close to the identity matrix.
This, in turn, makes most of the singular values of the matrix $PA\sqrt{W}$
close to one. However for very large values of \emph{$n-p$} the smallest
singular values can be less than the floating-point precision (about
$10^{-16}$ for 64-bit arithmetic) which will corrupt the calculated
singular values with roundoff error. To mitigate this problem, if
a calculated singular value is less than a particular threshold ,
$\delta$, then the inverse of the calculated singular value in Equation
8.26 is set to zero. An appropriate value of the parameter $\delta$
depends on the floating-point precision. For 64-bit arithmetic a value
of $10^{-15}$ gives satisfactory results.\smallskip{}

The procedure of setting the inverses of calculated singular values
less than a particular threshold to zero is called truncated SVD regularization.
It can be regarded as a weighting scheme where the weights of the
calculated singular value inverses change from one to zero at the
threshold. An alternative to this procedure, Tikhonov SVD regularization,
provides a smoother transition in the weights over a range of calculated
singular values in the vicinity of the threshold. An analysis of the
distribution of calculated singular values of the matrix $PA\sqrt{W}$
suggests that the transition to calculated singular values corrupted
by roundoff error is relatively sharp. Therefore truncated SVD regularization
is better suited for the discretization compensation algorithm.

\smallskip{}
When the discretization compensation algorithm is applied to the coefficients
of the Grace filter the resultant filter will be referred to as the
adjusted Grace filter. This filter is implemented in the program GraceA
in Appendix 2. (Much of the work in this study was done using Mathcad
software. Programs written in Mathcad worksheets look like pseudocode
by design. So the programs in the appendices are adapted from screenshots
from Mathcad worksheets.)\newpage{}

\section{The Performance of the Adjusted Grace Filter}

For the purpose of this section the \emph{k}th even order derivative
of $\mathrm{h}(f,c)$, evaluated at $\phi=0$ (Equation 7.5) will
be transformed by a multiplication of a function of \emph{m, p} and
\emph{k}, analogous to Equation 5.2. The transformed derivative will
be denoted by $\mathrm{d}_{k}^{\prime}(m,p,c)$ which  is given by:

\begin{equation}
\mathrm{d}^{\prime}(m,p,c,k)=\dfrac{4^{p+k}}{(m\pi)^{2k}}\left.\frac{d^{2k}\mathrm{h(\mathit{f,c})}}{df^{2k}}\right|_{f=0}=(-1)^{k}\,4^{p+k}\stackrel[i=-m]{m}{\sum}c_{i}\,x_{i}^{2k}
\end{equation}

The following equation applies for the coefficients of the Grace filter,
$c_{G}$, and the adjusted Grace filter, $c_{A}$:

\begin{equation}
\underset{m\rightarrow\infty}{\lim}\mathrm{d}^{\prime}(m,p,c_{G},k)=\underset{m\rightarrow\infty}{\lim}\mathrm{d}^{\prime}(m,p,c_{A},k)=\mathrm{d}(n,p,k)
\end{equation}

where $\mathrm{d}(n,p,k)$ is the transformed \emph{k}th even derivative
of the Fourier transform, $g(\phi)$, at $\phi=0$ given by Equation
5.2. The second and third columns of Table 9.1 give the values of
$\mathrm{d}^{\prime}(m,p,c,k)$ for the coefficients of the Grace
filter ($c_{G}$) and the adjusted Grace filter ($c_{A}$) for $m=15$,
$n=10$, $p=5$, and $k=1,\ldots\,12$. In this case the number of
zero-valued even derivatives of the Fourier transform of the Grace
function at $\phi=0$ is $z=n-p-1=4.$ The fourth column gives the
integer values of $\mathrm{d}(n,p,k)$ for $n=10$ and $p=5$.
\begin{center}
Table 9.1: Transformed even derivatives of the frequency response
at zero frequency.
\par\end{center}

\[
\begin{array}{rrrrr}
k &  & \mathrm{d}^{\prime}(m,p,c_{G},k) & \mathrm{d}^{\prime}(m,p,c_{A},k) & \mathrm{d}(n,p,k)\\
\\
1 &  & -0.000\quad & 0\quad & 0\quad\\
2 &  & 0.002\quad & 0\quad & 0\quad\\
3 &  & -0.010\quad & 0\quad & 0\quad\\
4 &  & 0.050\quad & 0\quad & 0\quad\\
5 &  & -1.223\quad & -1.004\quad & -1\quad\\
6 &  & 0.934\quad & 0.038\quad & 0\quad\\
7 &  & 6.285\quad & 9.775\quad & 10\quad\\
8 &  & -25.793\quad & -38.981\quad & -40\quad\\
9 &  & 42.435\quad & 91.256\quad & 95\quad\\
10 &  & 68.937\quad & -109.208\quad & -120\quad\\
11 &  & -871.572\quad & -228.215\quad & -210\quad\\
12 &  & 4460.556\quad & 2154.694\quad & 2200\quad
\end{array}
\]

Table 9.1 illustrates that in this case the discretization compensation
algorithm does more than just zero out the first \emph{z} even derivatives
of the frequency response at zero frequency. A row-wise comparison
of the columns shows that the algorithm also has the effect of bringing
the higher order even derivatives of the frequency response at zero
frequency more into agreement with the corresponding even derivatives
of the Fourier transform of the corresponding Grace function.\smallskip{}

Adjusting the filter coefficients to zero out the first $z=n-p-1$
even derivatives of the frequency response evaluated at zero frequency
flattens the frequency response in the neighbourhood of $f=0$. However
there remains the possibility that the discretization compensation
algorithm could suppress the amplitude of ripples near zero frequency
while amplifying the amplitude of ripples at higher frequencies. The
general issue of the monotonicity of the pass band frequency response
for the adjusted Grace filter will now be addressed.

\smallskip{}
In Section 6 the Fourier transform of the Grace function was transformed
by Equation 6.1 to highlight the nature of the Fourier transform in
the pass band. The frequency response of the adjusted Grace filter,
$\mathrm{h}(f,m,n,p)$, can be similarly transformed to highlight
the nature of the frequency response in the pass band:

\begin{equation}
\mathrm{f}^{\prime}(f,m,n,p)=\left[\,1-\mathrm{h}(f,m,n,p)\,\right]^{\tfrac{1}{2(n-p)}}
\end{equation}

\newpage The transformed frequency response of the adjusted Grace
filter is plotted in Figure 9.1 for $m=15$, $n=10$ and $p=0,5,6,7,8,9$.
The frequency has been scaled by dividing the frequency by the reference
frequency, $f_{r}$, given by Equation 7.7:

\begin{equation}
\phi=\dfrac{f}{f_{r}}
\end{equation}

\begin{center}
Figure 9.1: Plot of $\mathrm{f}^{\prime}(\phi,m,n,p)$ for $m=15$,
$n=10$ and $p=0,5,6,7,8,9$
\par\end{center}

\begin{center}
\includegraphics[scale=0.6]{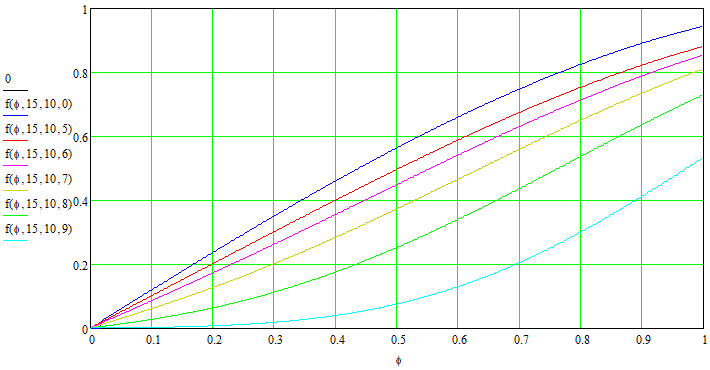}
\par\end{center}

Figure 6.1 shows the transformed Fourier transform of the Grace function
for the same values of \emph{n} and \emph{p}. The corresponding curves
in Figures 6.1 and 9.1 are virtually coincident over the entire pass
band.\smallskip{}

The close agreement in this particular case motivated a systematic
comparison between the transformed frequency response of the adjusted
Grace filter, $\mathrm{f}^{\prime}(\phi,m,n,p)$, and the transformed
Fourier transform of the corresponding Grace function, $\mathrm{f}(\phi,n,p)$.
Ripple in the pass band is most likely to manifest at smaller values
of the filter parameters where the adjustments made to the filter
coefficients are the largest. Therefore values of \emph{m} from 3
to 50, values of \emph{n} from 2 to $m-1$ and values of \emph{p}
from 0 to $n-2$ were considered. For each of these 19600 cases $\mathrm{f}^{\prime}(\phi,m,n,p)$
and $\mathrm{f}(\phi,n,p)$ were calculated at 1000 equally spaced
points over the frequency interval $\left[0,f_{r}\right]$ or $\left[0,1\right]$
on the $\phi$ scale. The metric used for the comparison in each case,
$\Delta\mathrm{f}$, is the root mean square (rms) of the differences
between the \emph{q} sets of points, i.e.:

\begin{equation}
\Delta\mathrm{f}=\sqrt{\dfrac{1}{q}\stackrel[i=1]{q}{\sum}\left(\mathrm{f}(\phi_{i},n,p)-\mathrm{f}^{\prime}(\phi_{i},m,n,p)\right)^{2}}
\end{equation}

The square of this metric is nearly equivalent to numerically evaluating
the integral of the squared difference between the two functions over
the interval $\left[0,1\right]$ using the trapezoidal rule with \emph{q}
points. For the 19600 cases considered the largest rms difference
was 0.019 for a value of \emph{p} of zero. In 96\% of the cases where
$p>3$ the rms difference was less than 0.001. This shows that the
discretization compensation algorithm has the unintended but beneficial
effect of bringing the filter frequency response into close agreement
with the Fourier transform of the Grace function over the entire pass
band. This property would not be expected of an ad hoc algorithm and
it suggests that the discretization compensation algorithm may be
uniquely well-suited for this application to the Grace filter. Why
this might be so merits further investigation.\newpage{}

\section*{References}

{[}1{]} Savitzky, A. and Golay M.J.E. ``Smoothing and differentiation
of data by simplified least-squares

\qquad{}procedures'' \emph{Anal. Chem.} 36 (1964) 1627-39

{[}2{]} Schafer, R.W. ``What is a Savitzky-Golay filter?'' \emph{IEEE
Signal Process. Mag}. July 2011, 111-117

{[}3{]} Sloane, N.J.A. The On-line Encyclopedia of Integer Sequences.
http://oeis.org

\section*{Acknowledgments}

The usefulness of David H. Bailey's MPFUN arbitrary precision Fortran
function library funded by NASA Ames and the helpful assistance of
Mahdi S. Hosseini of the University of New Brunswick are gratefully
acknowledged. Ad majorem Dei gloriam.

\section*{Contact}

The author can be contacted at pwoxby@telus.net

\section*{Appendix 1: The Grace Filter Coefficients Program}

The program Grace is based on Equations 1.3, 3.1 and 7.1. It departs
from the text in that the range of the coefficient subscripts, \emph{i},
is $\left[1,...\,2m-1\right]$ instead of $\left[-m,...\,m\right]$.
The zero-valued coefficients at the ends of the interval are not calculated.
The subscripts on \emph{u} and \emph{v} (Equation 1.3) used in the
text for clarity are superfluous and have been dropped.\smallskip{}

\includegraphics[scale=0.6]{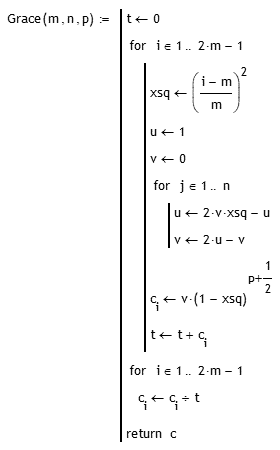}

\newpage{}

\section*{Appendix 2: The Adjusted Grace Filter Program}

The first part of the program GraceA, up to the SVD reference, is
based on Equations 8.21, 8.23 and 8.24. In the program the letters
rt in the names of variables stand for square root. In the loop where
the Chebyshev polynomials are calculated only the values of the even-order
polynomials are stored in the variable ue. The second part of the
program GraceA, starting with the SVD reference, is based on Equations
8.25, 8.26 and 8.27. The argument parameter $\delta$ reduces the
sensitivity of the SVD algorithm to potentially ill-conditioned matrices.
For 64-bit arithmetic a value for $\delta$ of $10^{-15}$ gives satisfactory
results. The range of the subscripts is either $\left[1,...\,2m-1\right]$
or $\left[1,...\,z+1\right]$, depending on the context.

\includegraphics[scale=0.6]{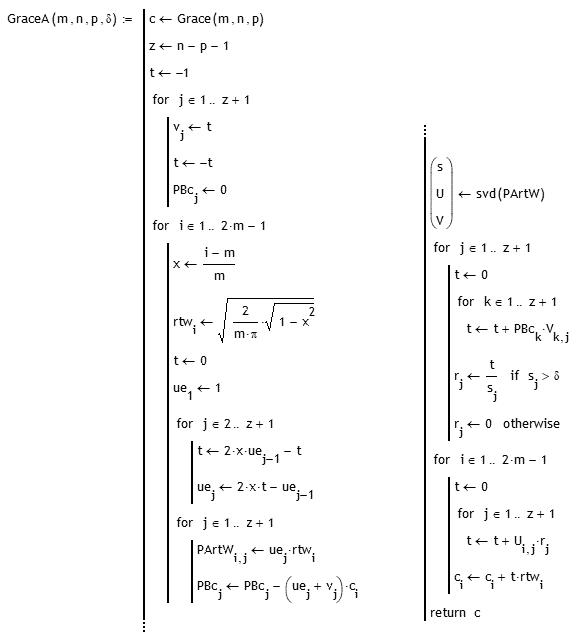}

\newpage{}

\section*{Appendix 3: More on the Grace Function Fourier Transform}

The results shown in Figures 6.1 and 6.2 suggest that $\mathrm{f^{\infty}}(\phi)$
(Equation 6.7) is an upper bound on $\mathrm{f}(\phi,n,p)$ (Equation
6.1). However the results shown in Figure 6.3 do support this conjecture.
This anomaly merits an explanation therefore a more detailed analysis
of the behaviour of the Fourier transform of the Grace function is
presented here.\smallskip{}

The results in Figures 6.1, 6.2 and 6.3 were calculated using Equations
5.10 and 6.1 for $\mathrm{f}(\phi,n,p)$. However the nature of the
behaviour of the curves in these figures is better understood by examining
the two leading coefficients, $r_{1}(n,p)$ and $r_{2}(n,p)$, of
Equation 6.2.

\[
\qquad\qquad\mathrm{f}(\phi,n,p)=\stackrel[i=1]{\infty}{\sum}r_{i}(n,p)\,\phi^{2i-1}\qquad\qquad\qquad(6.2)
\]

The values of the coefficients $r_{1}(n,p)$ and $r_{2}(n,p)$ are
given by Equations 6.3 and 6.4. In the following plots the six values
of the parameter \emph{n} range from 32 to 1024 by factors of two.
Given this relatively large range in the values of \emph{n} the values
of the parameter \emph{p} are expressed as the ratio of \emph{p} to
\emph{n} which ranges from zero to one. The notation $r_{i}(n,p)$
in the text is expressed in the following plots as $ri(n,p\_to\_n)$
where $p\_to\_n$ is the ratio of \emph{p} to \emph{n}. \emph{p} is
a discrete parameter but the intervals between the discrete values
of \emph{p} are interpolated in the plots for clarity. The six curves
for $r_{1}(n,p)$ are shown in Figure A3.1a:
\begin{center}
Figure A3.1a: $r_{1}(n,p)$ vs $\nicefrac{p}{n}$ for $n=32,64,128,256,512,1024$
\par\end{center}

\begin{center}
\includegraphics[scale=0.6]{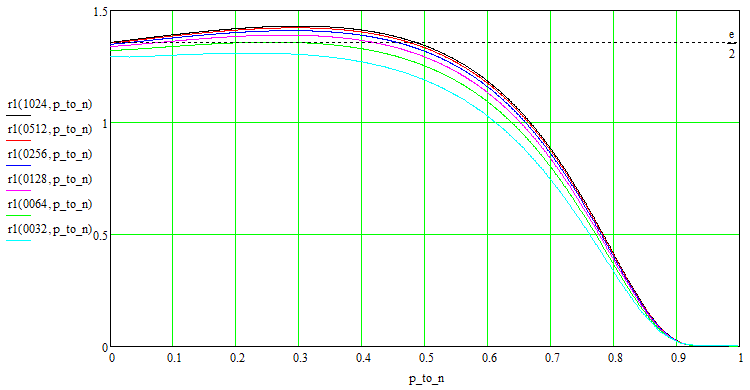}
\par\end{center}

In this figure the limiting value of $r_{1}(n,p)$ as \emph{n} approaches
infinity for a fixed value of \emph{p}, i.e., $r_{1}^{\infty}=\nicefrac{e}{2}$
(Equation 6.6), is indicated by the dashed line.\newpage The behaviour
of the six curves near the limiting value of $\nicefrac{e}{2}$ is
of particular interest. Figure A3.1b shows a closeup view of this
part of Figure A3.1a:
\begin{center}
Figure A3.1b: Closeup view of Figure A3.1a near $r_{1}(n,p)=\nicefrac{e}{2}$:
\par\end{center}

\begin{center}
\includegraphics[scale=0.6]{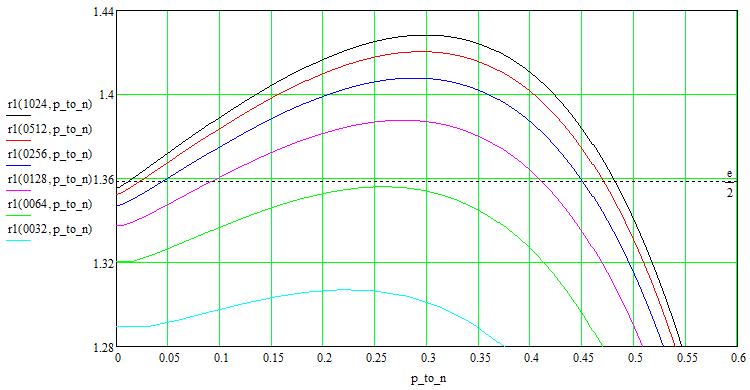}
\par\end{center}

As \emph{n} approaches infinity for constant \emph{p}, the ratio \emph{$\nicefrac{p}{n}$}
approaches zero. In this figure the trend in the values of $r_{1}(n,p)$
for $\nicefrac{p}{n}=0$ as the values of \emph{n} increase is consistent
with the limiting value, $r_{1}^{\infty}=\nicefrac{e}{2}$. But the
simple fact that values of $r_{1}(n,p)$ can exceed $r_{1}^{\infty}=\nicefrac{e}{2}$
for values of $\nicefrac{p}{n}>0$ is sufficient to disprove the conjecture
that $\mathrm{f_{g}^{\infty}}(\phi)$ (Equation 6.7) is an upper bound
on $\mathrm{f}(\phi,n,p)$ (Equation 6.1).\smallskip{}

The six curves for $r_{2}(n,p)$ are shown in Figure A3.2a:
\begin{center}
Figure A3.2a: $r_{2}(n,p)$ vs $\nicefrac{p}{n}$ for $n=32,64,128,256,512,1024$
\par\end{center}

\begin{center}
\includegraphics[scale=0.6]{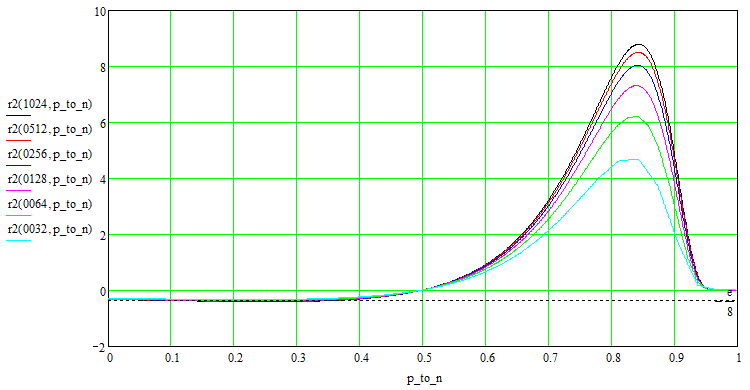}
\par\end{center}

In this figure the limiting value of $r_{2}(n,p)$ as \emph{n} approaches
infinity for constant \emph{p}, i.e., $r_{2}^{\infty}=-\nicefrac{e}{8}$
(Equation 6.8), is indicated by the dashed line. The behaviour of
the six curves near the limiting value of $-\nicefrac{e}{8}$ is of
particular interest. Figure A3.1b shows a closeup view of this part
of Figure A3.1a:
\begin{center}
Figure A3.2b: Closeup view of Figure A3.2a near $r_{2}(n,p)=-\nicefrac{e}{8}$:
\par\end{center}

\begin{center}
\includegraphics[scale=0.6]{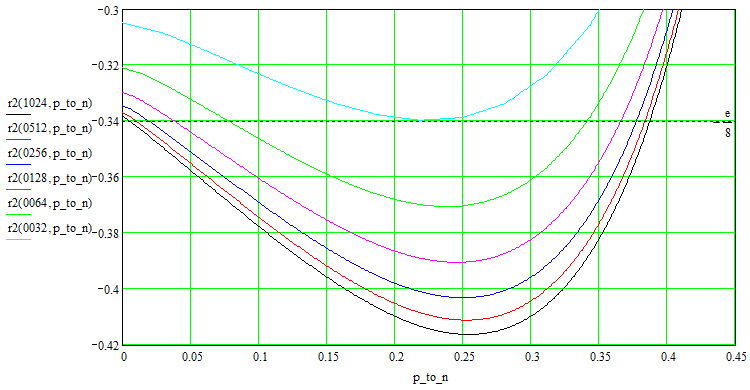}
\par\end{center}

In this figure the trend in the values of $r_{2}(n,p)$ for $\nicefrac{p}{n}=0$
as the values of \emph{n} increase is consistent with the limiting
value, $r_{2}^{\infty}=-\nicefrac{e}{8}$. Collectively these plots
show why the conjecture that $\mathrm{f^{\infty}}(\phi)$ is an upper
bound on $\mathrm{f}(\phi,n,p)$ is false. The conjecture is based
on the simple behaviour of the plots for $\nicefrac{p}{n}=0$ but
this behaviour is not representative of the significantly more complex
general behaviour of the plots.\smallskip{}

Another area of interest in Figure A3.2a is the region near $\nicefrac{p}{n}=0.5$.
It appears that the six values of $r_{2}(n,p)$ are all zero for $\nicefrac{p}{n}=0.5$
and this is, in fact, the case. The denominator of Equation A3.2 for
$r_{2}(n,p)$ contains the factor $n-2p$. Therefore $r_{2}(n,p)=0$
for even \emph{n} and $p=\nicefrac{n}{2}$. The sign of $n-2p$ determines
whether $\mathrm{f_{g}}(\phi,n,p)$ is convex or concave at $\phi=0$.
$\mathrm{f_{g}}(\phi,n,p)$ at $\phi=0$ is concave for $p<\nicefrac{n}{2}$
and convex for $p>\nicefrac{n}{2}$. The value of $p=\nicefrac{n}{2}$
conveniently divides the general behaviour of the Fourier transform
of the Grace function into two distinct regions. 
\end{document}